\documentclass{article}
\usepackage[utf8]{inputenc}

\usepackage{color}
\usepackage{graphicx}
\usepackage{tikz}

\usepackage{mathtools}
\usepackage{multirow}
\usepackage{url}
\usepackage{natbib}

\title{Correlated evolution: models and methods.}
\author{G Achaz$^{1,2}$, JY Dutheil$^{3,4}$\\
\date{\small{1: UMR7206 Eco-Anthropologie, Université de Paris--CNRS--MNHN, Paris \\
  2: UMR7241 Centre Interdisciplinaire de Recherche en Biologie, Collège de France--CNRS--INSERM, Paris \\
  3: AG Molecular Systems Evolution -- Max Planck Institute for Evolutionary Biology, Plön, Germany \\
  4: Institute des Sciences de l'Evolution - Montpellier, Université Montpellier--CNRS--EPHE--IRD, Montpellier, France\\
  \texttt{Corresponding author: guillaume.achaz@mnhn.fr}}}}

\begin{document}

\maketitle

\newpage
\tableofcontents

\newpage

\section{Introduction}
As soon as the word \textit{co-evolution} is dropped, the attention of evolutionary biologists rises. Some pictures illuminating patterns of co-evolution between species such the joint evolution of the beak of the beautiful pollinating hummingbirds and its associated flowers \citep{kay_rapid_2005} or the terrifying arm races between ferocious pathogens and their hosts \citep{woolhouse_biological_2002}; soon after, many will think about Alice running on the spot with the Red Queen through the looking glass \citep{van_valen_new_1973}. Our fascination about complex evolving entities that are in constant dynamical interactions often lies at the cross-road between Ecology and Evolution, both being arguably parts of the field coined Population Biology \citep{lewontin_building_2004}.

There is, however, another common usage of the term \textit{co-evolution}, popular among molecular evolutionary biologists, where the co-evolving entities are not interacting species of an ecological community, but traits of the same individual, species or genome. This chapter reviews a selection of models and methods related to this second meaning. We use equally the terms \textit{co-evolution} and \textit{correlated evolution} to name the joint evolution of two or more traits; \textit{trait} refers to \textit{any} character of individuals, ranging from  morphological traits (e.g. height) to genetic variant (e.g. an allele at a locus). We are naturally inclined to accept patterns of co-evolution between the height and the body mass of individuals or, at the molecular level,  between paired nucleotides of the same RNA molecule. We show in this review how, once abstracted, the models and methods that have been used to characterize the correlated evolution between morphological traits or between nucleotides share several common features. They  are designed in the same general framework and only differ by details of their implementation. 

To demonstrate correlated evolution between two traits, one has to reject the statistical independence between the two  processes that govern the evolution of the two traits. To do so, one can look at the state of the two traits in many individuals, species or genomes and characterize the co-occurrence or co-abundance of the two traits among the cohort. Alternatively, one can compare the evolutionary paths of the two trait states and scan for signs of co-variations. The hope is that the patterns of correlated evolution is caused by biologically relevant interactions between the two traits. However, as for any pattern of correlation, the non-independence can be indirectly caused by correlations to a third hidden variable. For instance, an environmental variable may drive synchronised changes in several traits that are not interacting directly. As always, \textit{correlation does not imply causation}. 

\begin{figure}[ht]
    \centering
    \includegraphics[width=\textwidth]{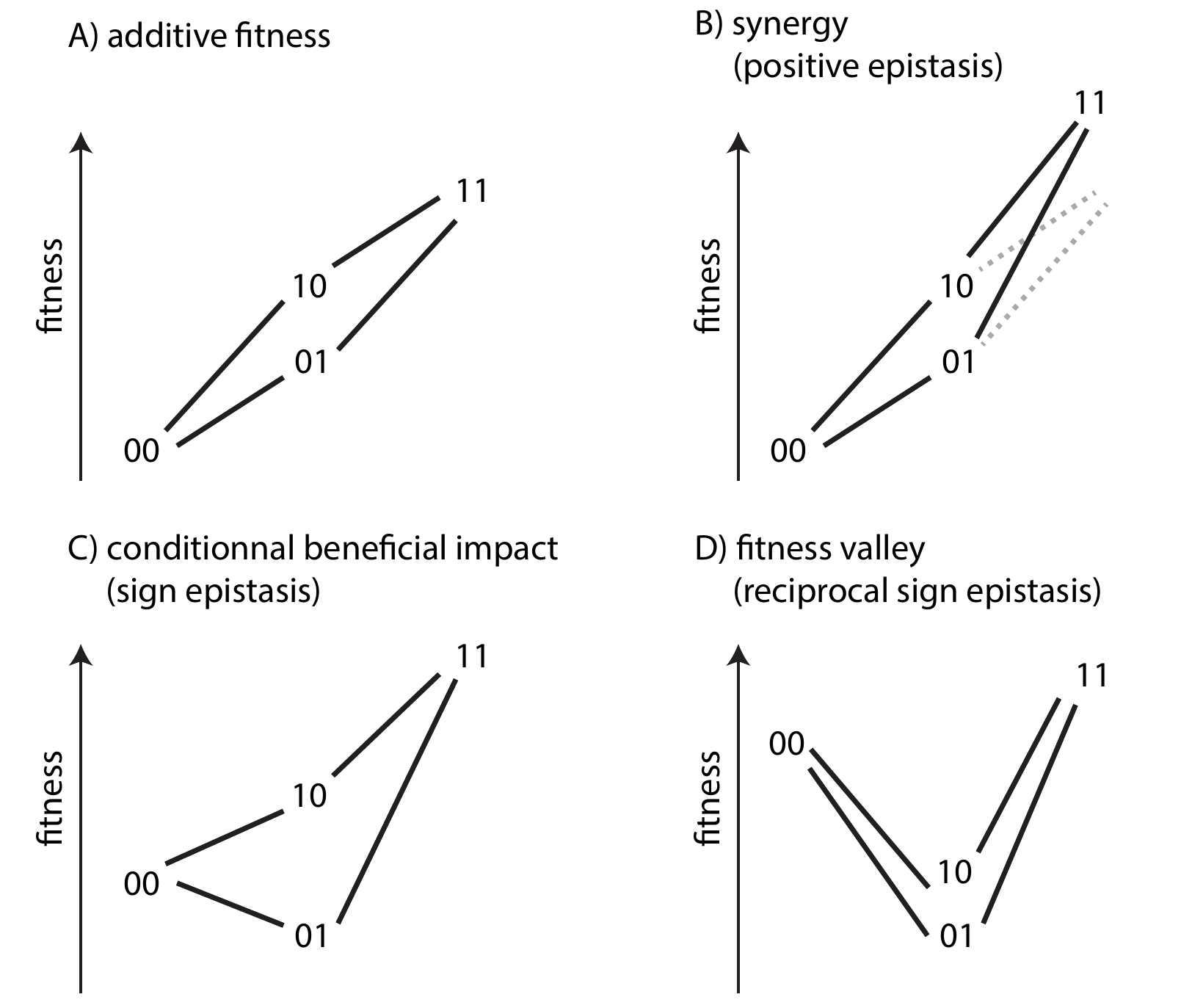}
    \caption{\small{Four examples of fitness landscapes with 2 binary traits, resulting in four possible combinations. The height of a combination represents its fitness. A) When the fitness difference of the double mutant compared to the reference, is simply the sum of each mutation, there is no dependency between the two mutations. The landscape is named additive and the two traits evolve independently. B) When the fitness improvement of the double mutant is more than the additive case, there is a synergistic effect of both mutations. C) The mutation of the second trait is deleterious in the reference ($00 \to 01$) but beneficial once the first trait has mutated ($10 \to 11$). D) Both traits are deleterious when mutated on their own but compensate each other in the double mutant.}}
    \label{figure:fl}
\end{figure}

When the evolution of the two traits are actually directly interacting, it is usually the case that their joint mutations have a different impact on fitness than what is expected by the mutation of each one. For example, they can have a synergistic effect when mutating both traits together result in a larger fitness improvement than the sum of each mutation. At the molecular level, interactions between loci, a phenomenon known as epistasis, were the subject of long-standing interests for geneticists \citep{phillips_epistasis_2008}. In its simplest form, let's consider two binary traits. An individual can then have 4 possible trait states: 00, 01, 00, 11. Each of the combination can be associated to a fitness value. The map of all possible combinations of trait states to a fitness value is known as a fitness landscape \citep{wright_roles_1932}. For the simple case of 2 traits with two states, the fitness landscape can be easily sketched as in Figure \ref{figure:fl}.

Depending on which fitness landscape both traits are evolving in, one expects different patterns of correlated evolution. Probably the strongest patterns of co-evolution would arise when only the double mutant is not associated to a fitness loss (Figure \ref{figure:fl}d). It corresponds to cases of compensation where the second mutation will compensate, or even over-compensate, the first one. In such a case, we would only observe individuals with trait combinations 00 or 11, and more rarely the intermediate states. Similarly, we expect to only witness synchronized mutations in the evolutionary paths of the two traits, where one mutation in a trait is immediately followed by a mutation in the other. Fitness landscapes have fascinated evolutionary biologists (among others) for almost a century and are still nowadays an object of vivid interest \citep{achaz_reproducibility_2014, visser_empirical_2014, yi_adaptive_2019}.

In this chapter, we will first explore models of correlated evolution between continuous and discrete traits (section 1). These models are historically used to characterize how traits are co-varying in the natural world as species evolve through time. We will then continue our journey with models of correlated evolution within genomes (section 2), that have many times been successfully applied to understand co-variations within the same molecule (RNA or proteins), but also at a larger scale to assess co-evolution between different genes or even between regulatory elements and genes. Finally, we will show how standard classical genetics can be rethought in terms of correlated evolution (section 3) between phenotypic trait(s) and molecular trait(s) in the genome.

In this chapter, we review the broad corpus of models and methods that have been developed to study coevolution. We more particularly explain the motivations and basic principles of a few chosen models and their related methodologies. We hope that this review will provide to the readers an original overview of the field of \textit{correlated evolution} by integrating evolution, genetics, structural biology, statistical physics, modelling and statistics. 

\section{Correlated evolution between traits}
\subsection{Species are not independent}

Co-evolution is expected to result in correlations of character states when measured in several species. Comparative analysis, therefore, constitutes a natural method of investigation to exhibit coevolving traits. The simple comparative analysis of observed states, however, could be misleading as biological species are not statistically independent. Quite on the contrary, as all species share ancestry, the quest of correlated evolution requires dedicated methods. Ignoring the phylogeny results in biases when simple counting methods are applied to species traits. We illustrate this phenomenon, sometimes termed ``phylogenetic inertia'' \citep{harvey_comparative_1991}, with a correlation test.

\begin{figure}[t]
    \centering
    \includegraphics[width=\textwidth]{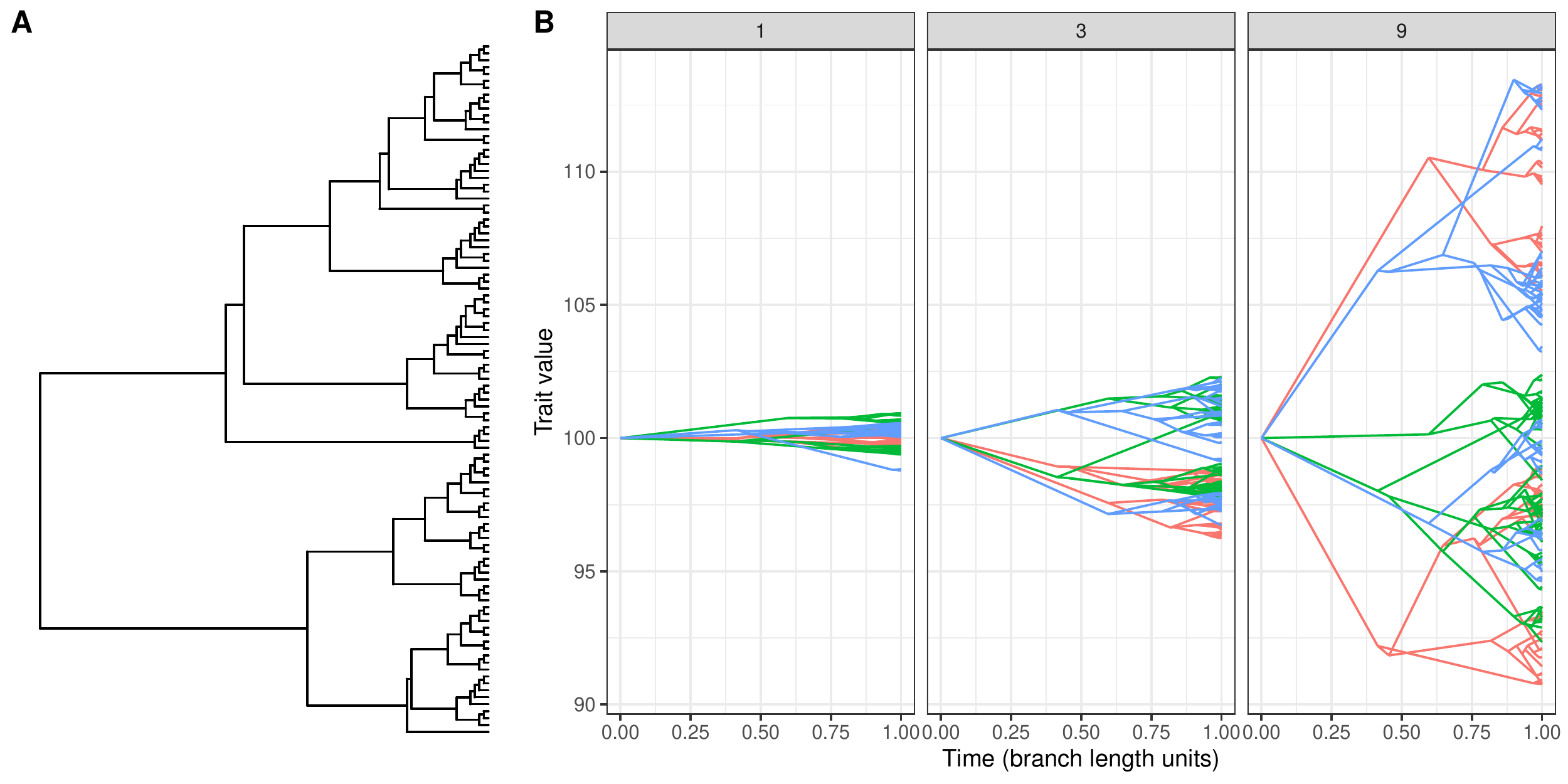}
    \caption{\small{Evolution of a continuous trait under a Brownian motion along a phylogeny. A) A 100 species random phylogeny used for simulation. B) Simulations under the phylogeny depicted in A) using a Brownian motion with variance equal to 1, 3 or 9. Three independent replicates are displayed with three distinct colors in each case. Note the emergent co-variation in the traits, that is only due to the shared phylogeny.}}
    \label{fig:brownian.tree}
\end{figure}

We must first acknowledge the history partially shared by species, that is classically represented by a phylogenetic tree. We will, therefore, account for the phylogenetic relationships between the species by modelling the evolution of traits along the tree. The most simple model of continuous trait evolution was introduced by \citet{felsenstein_maximum-likelihood_1973}. Under this model, a trait $Y$ mutation is only described by its intrinsic variance $\sigma^2_Y$. The evolution of the trait is modelled as a Brownian motion: the distribution $D_Y(t)$ of the evolved trait after a certain time $t$ is a normal distribution with mean $y_0$ (trait at time 0) and a variance equal to $\sigma^2_Y \times t$.
To simulate the evolution of a trait along a tree, we can apply this procedure recursively on each branch of the tree, propagating the trait values down the tree.
We note $y_i$ the (ancestral) traits at node $i$ in the tree; $y_0$ is the ancestral trait state at the root (node 0) of the tree. We further note the branch lengths $\{t_i\}_{1 < i < n}$ between a node $i$ and its parent $p$ in a rooted tree with $n$ branches. Starting from the root of the tree, a value for each descendant node $i$ is then drawn randomly from the distribution
\begin{equation}
    D_Y(t_i) = {\cal N}(y_p, \sigma^2_Y \times t_i),
\end{equation}
where $y_p$ is the value of the trait at the parent node. Figure \ref{fig:brownian.tree} shows an example of three traits simulated with Brownian motions with distinct intrinsic variances.

Traits can be then measured at the leaves of the tree, representing a sample of species with partially shared histories. Once the traits are measured in the different species, one can compute simple correlations between the trait values. Here we have chosen the Kendall's $\tau$ test because it is a non-parametric rank correlation test between two numerical variables, without any assumption regarding the distribution of the data. Like most correlation tests, it considers each pair of trait values (\textit{i.e.} a species) as an independent sampling. Under the null hypothesis of independence, if one draw randomly one set of pairs of values that have two independent distributions and compute their correlation, one has 5\% chances to obtain a P value below 0.05. More generally, if one computes the distribution of P values over many such samples, we obtain a uniform distribution over $[0,1]$.

When simulating two independent traits along the same phylogeny and assessing their correlation, we observe that the distribution of P values is skewed toward small values (Figure \ref{fig:pvaldist}). This means that the two traits appear intrinsically correlated, despite being independently simulated along the phylogeny.

This implies that the correlation between two traits in a sample of species is the sum of an intrinsic correlation and a phylogenetic correlation. As a consequence, there is a need for methods that specifically test for the significance of the intrinsic correlation, eliminating for the phylogenetic one. One possibility is to only sample species that are very distant and/or have emerged from a rapid radiation and exhibit a star-like  phylogeny. Another is to specifically incorporate the phylogeny in the method.

\subsection{The phylogenetically independent contrasts}

\begin{figure}[t]
    \centering
    \begin{tikzpicture}
    \draw(0,8) node[above]{1} -- node[left]{$t_1$} (0,5) -- (2,5) -- node[left]{$t_2$} (2,6) node[above]{2};
    \draw(1,5) node[above]{8} -- node[left]{$t_8$} (1,3) -- (3,3) -- node[left]{$t_3$} (3,7) node[above]{3};
    \draw(4,8) node[above]{4} -- node[left]{$t_4$} (4,6) -- (6,6) -- node[left]{$t_9$} (6,9) node[above]{5};
    \draw(7,7) node[above]{6} -- node[left]{$t_6$} (7,5) -- (9,5) -- node[left]{$t_7$} (9,7) node[above]{7};
    \draw(5,6) node[above]{10} -- node[left]{$t_{10}$} (5,4) -- (8,4) -- node[left]{$t_{11}$} (8,5) node[above]{11};
    \draw(2,3) node[above]{9} -- node[left]{$t_9$} (2,0) -- node[above]{13} (6.5,0) -- node[left]{$t_{12}$} (6.5,4) node[above]{12};
    \end{tikzpicture}
    \caption{Example rooted phylogeny with node labels and branch lengths}
    \label{fig:pictree}
\end{figure}
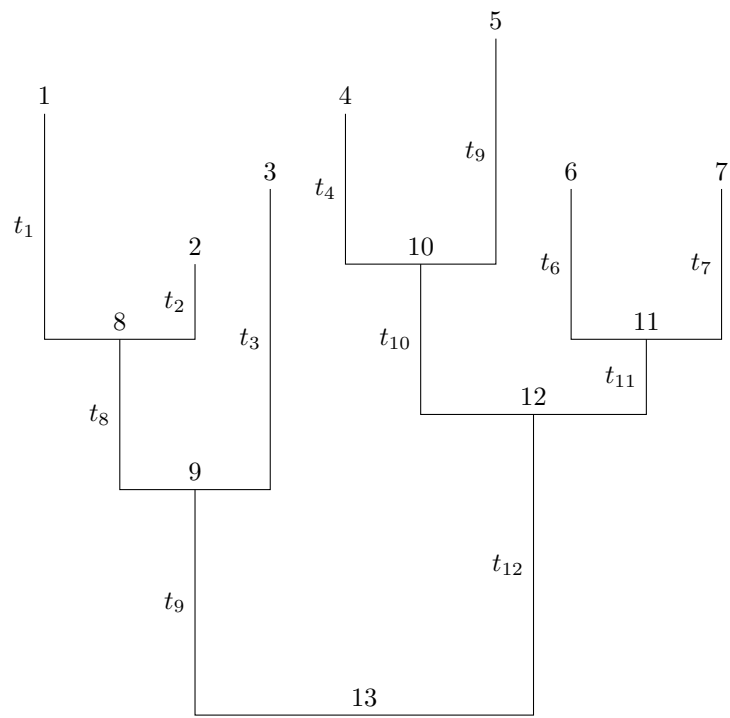
Using properties of the Brownian motion, \citet{felsenstein_phylogenies_1985} introduced one of the first comparative method for continuous traits, named the phylogenetic independent contrasts (PIC). The idea behind the contrasts method is a transformation of the input variables, so that the values of each transformed variable become statistically independent. Considering a given phylogenetic tree, it occurs that the evolution of a trait $y$ along two branches leading to a pair of node siblings $i$ and $j$ (e.g. 1 and 2) sharing the same parent node $k$ (e.g. 8) is independent. As a result, the differences $y_1 - y_2$, $y_4 - y_5$, $y_6 - y_7$, $y_8-y_3$, $y_{10} - y_{11}$ and $y_9 - y_{12}$ are all statistically independent (Figure \ref{fig:pictree}). Following this logic it is possible to transform a vector of $n$ trait values $\{y_i\}_{1 \leq i \leq n}$ into $n-1$ phylogenetically independent contrasts $\gamma_{ij} = y_i - y_j$.

Assuming a Brownian motion, the expectation and variance of the contrasts are:
\begin{eqnarray}
E(\gamma_{ij}) &=& 0 \\
V(\gamma_{ij}) &=& \sigma^2_Y \cdot (t_i + t_j).
\end{eqnarray}
To compute contrasts at internal nodes, we need to reconstruct ancestral trait states. The ancestral trait value $y_k$ for any ancestral node $k$ with two daughter branches $i$ and $j$ is obtained by taking the weighted mean of the trait values at the direct descendant nodes:
\begin{equation}\label{eqn:brownianancestor}
    y_k = \frac{\frac{1}{t_i}y_i - \frac{1}{t_j}y_j}{\frac{1}{t_i} + \frac{1}{t_j}}.
\end{equation}
We can therefore compute all contrasts and ancestral traits recursively along the tree, collapsing nodes after computing their associated contrasts. For instance, using the tree depicted on Figure \ref{fig:pictree}, one computes first $\gamma_8 = y1 - y_2$ and $y_8 = \left({t_1^{-1}y_i - t_2^{-2}y_j}\right)/\left({t_1^{-1} + t_2^{-1}}\right)$ and then removes leaves 1 and 2. Branch length $t_8$ is then updated in order to accommodate the fact that $y_8$ was estimated and not observed, and has an additional estimation variance. This is equivalent to an additional time for $y_8$ to have evolved from $y_9$, which depends on branch lengths $t_1$ and $t_2$ \citep{felsenstein_phylogenies_1985}:
\begin{equation}
    t_8' = t_8 + \frac{t_1\cdot t_2}{t_1 + t_2}.
\end{equation}
$y_8$ and $t_8'$ are then used to compute $\gamma_9$ and $y_9$, using formula \ref{eqn:brownianancestor}.

\begin{figure}[t]
    \centering
    \includegraphics[width=\textwidth]{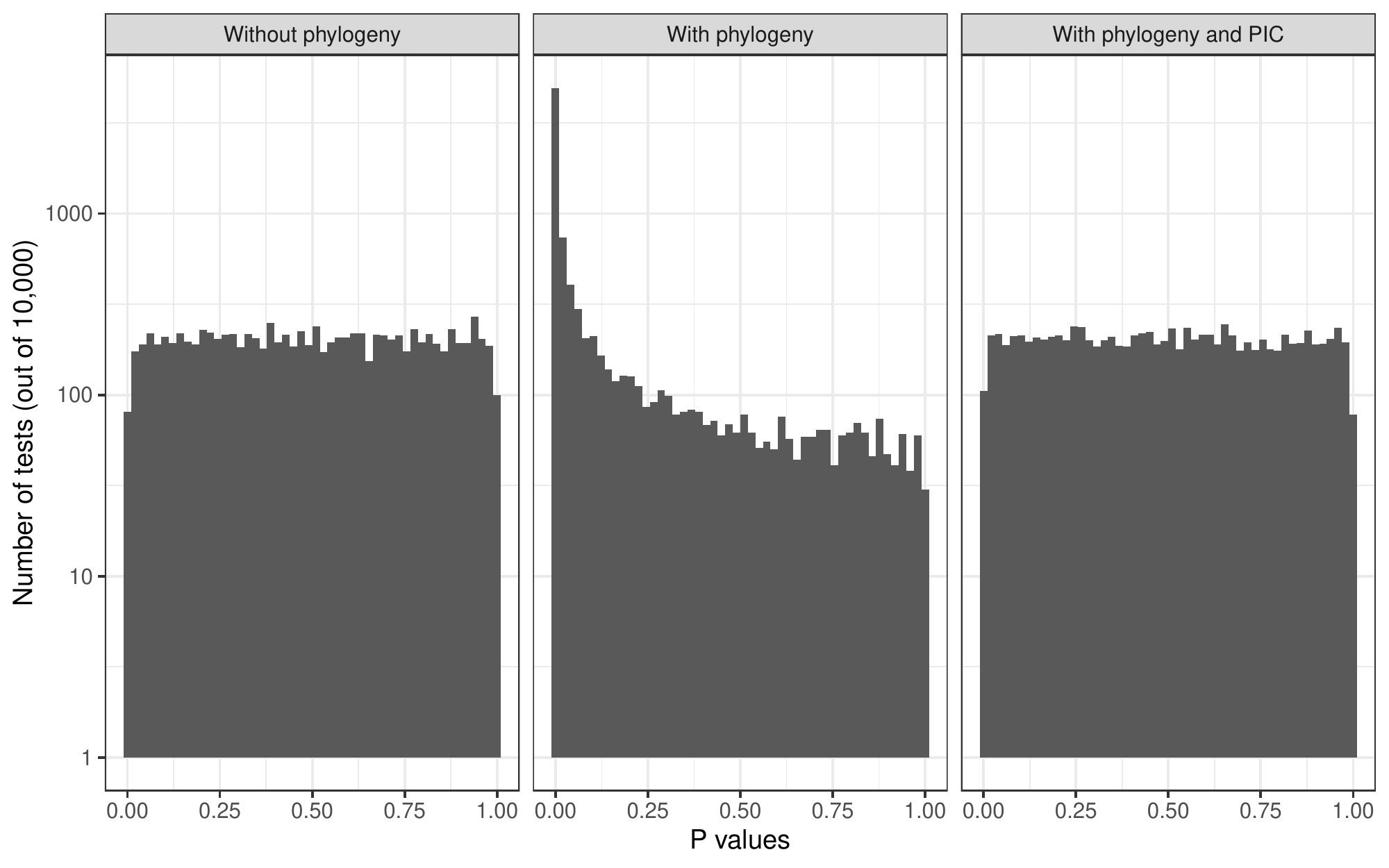}
    \caption{P value distribution of Kendall's correlation test on two continuous traits simulated independently. Left: without phylogeny or a star-like phylogeny (no shared history). Center: with phylogeny. Right: with phylogeny and correction using phylogenetic independent contrasts (PIC). 10,000 simulations were performed in each case, with a 100 pairs of points.
    \label{fig:pvaldist}}
\end{figure}

A simple method to test for the correlation between two continuous traits while accounting for the species phylogeny consists in first transforming each variable into independent contrasts, before testing for a correlation using standard classical tests. Figure \ref{fig:pvaldist} shows that this method successfully removes the phylogenetic correlation, at least when the true phylogeny is known and when the underlying trait evolves under a Brownian motion.

\subsection{Extending the linear model to account for phylogeny}

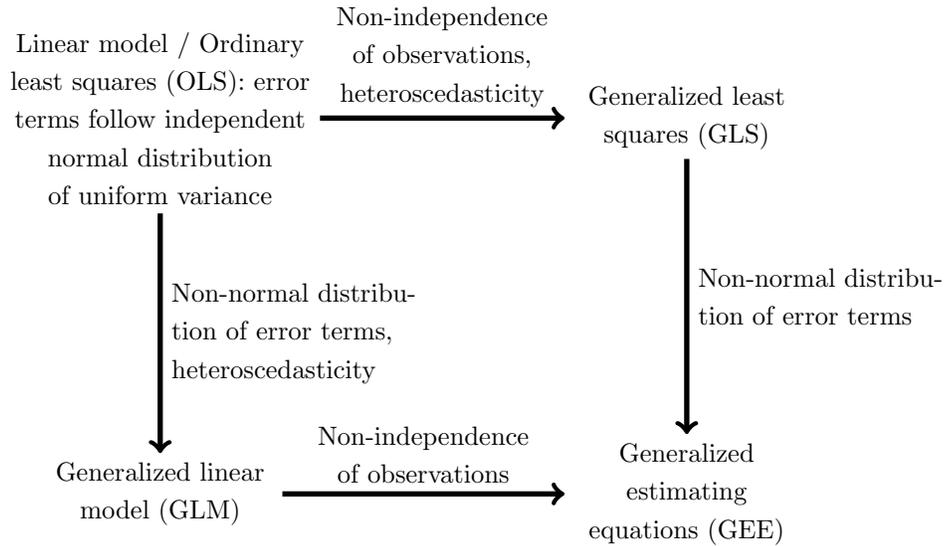
\begin{figure}[t]
    \centering
    \begin{tikzpicture}
    \draw(0,5) node (LM) [text width=4cm, align=center]{Linear model / Ordinary least squares (OLS): error terms follow independent normal distribution of uniform variance};
    \draw(0,0) node (GLM) [text width=3cm, align = center]{Generalized linear model (GLM)};
    \draw(7,0) node (GEE) [text width=3cm, align = center]{Generalized estimating equations (GEE)};
    \draw(7,5) node (GLS) [text width=3cm, align = center]{Generalized least squares (GLS)};
    \draw [->, line width = 2pt](LM) -- node[above, text width = 4cm, align = center]{Non-independence of observations, heteroscedasticity} (GLS);
    \draw [->, line width = 2pt](GLM) -- node[above, text width = 4cm, align = center]{Non-independence of observations} (GEE);
    \draw [->, line width = 2pt](LM) -- node[right, text width = 3.5cm, align = left]{Non-normal distribution of error terms, heteroscedasticity} (GLM);
    \draw [->, line width = 2pt](GLS) -- node[right, text width = 3.5cm, align = left]{Non-normal distribution of error terms} (GEE);
      \end{tikzpicture}
    \caption{The linear modeling framework and its extensions}
    \label{fig:linearmodel}
\end{figure}

The phylogenetic independent contrasts offer a simple way to compare different species traits. It proves, however, to be limited when testing hypotheses about trait evolution and environment variables, which are typically unknown for the ancestral species. As a consequence, these analyses are conducted within the linear modeling framework, where a given response variable $y$ with $n$ values $\left\{y_i\right\}_{1\leq i \leq n}$ is modeled as a linear combination of $m$ explanatory variables $\left\{x_j\right\}_{1\leq j \leq m}$ with corresponding values $\left\{x_{ji}\right\}_{1\leq j \leq m, 1\leq i \leq n}$:
\begin{eqnarray}
    y_i &=& \hat{y_i} + \epsilon_i\\
    \hat{y_i} &=& a_0 + \sum_{j = 1}^m a_j x_{j,i},
\end{eqnarray}
where $a_0$ is a constant (aka intercept) and the $\{a_j\}_{1\leq j\leq m}$ are the model coefficients. The $\left\{x_{ji}\right\}$ terms may denote observed variables or some of their interactions. For two measured traits, there can be up to three possible variables (the two traits and their pairwise interaction), for three measured traits, there can be be up to seven variables (the three observed traits, three possible pairwise interactions, and one triple interaction), etc.
In a typical evolutionary comparative analysis, $y$ would be a biological trait of interest (\textit{e.g.} body mass, age of sexual maturity, sexual dimorphism, etc) and the $x_j$ variables are either other biological variables such as life history traits, and/or environmental factors such as temperature, amount of precipitation, etc.

In the linear modeling framework, the model parameters $\{a_j\}$ are estimated using a maximum likelihood approach. The Gauss-Markov theorem states that, if the error terms $\{\epsilon_i\}_{1\leq i\leq n}$ are uncorrelated and with identical variance (homoscedasticity), the best unbiased linear estimators are the ordinary least square (OLS) estimators, that is, the values of $\{a_j\}$ for which $\sum_{i = 1}^n {(y_i - \hat{y_i})}^2 = \sum_{i = 1}^n \epsilon_i^2$ is minimal. Noting $Y = (y_i)_{1\leq i \leq n}$ as the response vector and $X = (x_{i,j})_{1\leq i\leq n, 1\leq j\leq m}$ the data matrix, such as
\begin{equation}
    X = \left(\begin{array}{cccccc}
         1 &  x_{1,1} & \cdots & x_{1,j} & \cdots & x_{1,m}\\
         \vdots & \vdots & \cdots & \vdots & \cdots & \vdots\\
         1 &  x_{i,1} & \cdots & x_{1,j} & \cdots & x_{i,m}\\
         \vdots & \vdots & \cdots & \vdots & \cdots & \vdots\\
         1 &  x_{n,1} & \cdots & x_{n,j} & \cdots & x_{n,m}\\
    \end{array}\right),
\end{equation}
and $A = (a_j)_{0\leq j\leq n}$, the OLS estimator $\hat{A}$ is obtained by the formula
\begin{equation}
    \hat{A}_{OLS} = (X^T X)^{-1} X^T Y,
\end{equation}
where $X^T$ is the transposed matrix of $X$. If the error terms are, in addition, normally distributed, then the OLS estimators are also the maximum likelihood estimators. 

The generalized linear model (GLM) framework extends the OLS estimates to a full class of models where the error terms follow distributions from the exponential family (Figure \ref{fig:linearmodel}). The GLM framework also includes heteroscedastic models, where the variance of the error term is not uniform. A strong assumption, however, remains: that the error terms are uncorrelated, which is at odds with the tree-based history of the biological data. 

Another extension, the generalised least square (GLS) framework, integrates a correlation structure into the ordinary least square estimators. Basically, it includes a variance-covariance matrix of individuals in the trait space, noted $V$. $V$ is a square matrix of size $n$, the sample size. The GLS estimates of $A$ are then simply given by the formula
\begin{equation}
    \hat{A}_{GLS} = (X^T V^{-1} X)^{-1} X^T V^{-1} Y.
\end{equation}
An interesting property of the GLS procedure is that the variance-covariance matrix, $V$, does not need to be fully specified, and free parameters can be estimated jointly with the regression coefficients within the maximum likelihood framework.

\begin{figure}
    \centering
    \begin{tikzpicture}
    \draw(0,5) node[above]{1} -- node[left]{$l_1$} (0.5,3) -- node[right]{$l_2$} (1,4) node[above]{2};
    \draw(0.5,3) -- node[left]{$l_6$} (1,2) -- node[right]{$l_3$} (2,4.5) node[above]{3};
    \draw(3,5) node[above]{4} -- node[left]{$l_4$} (3.5,3.5) -- node[right]{$l_5$} (4,4.5) node[above]{5};
    \draw(0.5,3) -- node[left]{$l_7$} (2,0) -- node[right]{$l_8$} (3.5,3.5);
    \end{tikzpicture}
    $$
    V_{brownian} = \sigma^2\left(
    \begin{array}{ccccc}
    l_1 + l_6 + l_7 & l_6 + l_7 & l_7 & 0 & 0\\
    l_6 + l_7 & l_2 + l_6 + l_7 & l_7 & 0 & 0\\
    l_7 & l_7 & l_3 + l_7 & 0 & 0 \\
    0 & 0 & 0 & l_4 + l_8 & l_8 \\
    0 & 0 &0 & l_8 & l_5 + l_8
    \end{array}
    \right)
    $$
    \caption{Variance-covariance matrix obtained from a rooted phylogeny with branch lengths, assuming a Brownian motion for the trait evolution. $\sigma^2$ is the intrinsic variance of the continuous trait.}
    \label{fig:vcvbrownian}
\end{figure}

Assuming a Brownian motion to model the evolution of continuous traits, the matrix of variances and covariances between species can be obtained directly from the branch lengths of the corresponding phylogenetic tree (Figure \ref{fig:vcvbrownian}): the variance for each species is given by the total branch lengths from the root of the tree, while the covariance between two species is equal to the total length between the root and the most recent common ancestor of the two species. \citet{grafen_phylogenetic_1989} was the first to propose the use of a GLS procedure as a way to account for phylogeny in the comparative analysis. The method was later extended by \citet{martins_estimating_1994} and \citet{martins_phylogenies_1997} and re-branded as 
``Phylogenetic Generalized Least Squares'' (PGLS). When a purely bifurcating tree is used with a Brownian motion model, the PGLS boils down to the simpler PIC method \citep{felsenstein_inferring_2003}. \citet{paradis_analysis_2002} later further extended the framework using generalized estimating equations (GEE, Figure \ref{fig:linearmodel}). A powerful property of the GLS and GEE frameworks is the possibility to test different models of evolution along a phylogeny, beyond the simple Brownian motion. \citet{blomberg_testing_2003}, for instance, extended the original Brownian motion to include heterogeneity in the rate of evolution.

Regardless of the model, there is a need for a phylogeny with estimated branch lengths. It is typically obtained from a different source, for instance molecular sequences. In many cases, the resulting branch lengths, which depict the evolutionary rates of particular genes, have \textit{a priori} little to do with the evolution of the trait under study. Branch lengths, however, cannot be estimated from the trait data as this would result in too many free parameters.
\citet{grafen_phylogenetic_1989} proposed a method to compute branch lengths for a given topology with a single parameter $\rho$. While the $\rho$ parameter does not have a particular biological interpretation, it adjusts the amount of treelike evolution: while $\rho = 0$ results in all internal branches to be equal to 0, effectively turning the phylogeny into a star tree with null covariances, $\rho = 1$ reduces all terminal branches to 0 and the data to simply two points. Estimating $\rho$ together will the regression parameters thus offers a ways to quantify the amount of phylogenetic correlation in the data, and to account for it when testing other effects. Finally, \citet{martins_phylogenies_1997} departed from the Brownian motion and considered a models where the covariance between two taxa depends on their phylogenetic distance $t_{ij}$:
\begin{equation}
V_{ij} = \gamma \exp(-\alpha\cdot t_{ij}).
\end{equation}
The matrix $V$ corresponding to the tree in figure \ref{fig:vcvbrownian} is:
\begin{equation}
   V_{Martins} = \gamma \exp \left[-\alpha\left(
    \begin{array}{ccccc}
    0 & l_{1,2} & l_{1,6,3} & l_{1,6,7,8,4} & l_{1,6,7,8,5}\\
    l_{2,1} & 0 & l_{2,6,3} & l_{2,6,7,8,4} & l_{2,6,7,8,5}\\
    l_{3,6,1} & l_{3,6,2} & 0 & l_{3,7,8,4} & l_{3,7,8,5} \\
    l_{4,8,7,6,1} & l_{4,8,7,6,2} & l_{4,8,7,3} & 0 & l_{4,5} \\
    l_{5,8,7,6,1} & l_{5,8,7,6,2} & l_{5,8,7,3} & l_{5,4} & 0
    \end{array}
    \right)\right]
\end{equation}
where $l_{x,y,z,\ldots} = l_x + l_y + l_z + l_{\ldots}$ and $\exp[M]$ denotes the element-wise exponential of a matrix $M$. The coefficients $\alpha$ and $\gamma$ can be interpreted in terms of constraining force and inter-specific variance of a model of stabilizing selection \citep{martins_phylogenies_1997}.

\subsection{Correlation between discrete traits}

Although methods for correlated evolution of discrete traits can be studied by the linear model framework, discrete traits coevolution has been the subject of specific developments. Building on the methodology of maximum parsimony, the transitions between trait states can be inferred and mapped on the branches of a phylogeny. In its simplest form, one first reconstructs the ancestral states at all internal nodes, and then counts one transition event on each branch where the states at the ancestral and descendant nodes are different (we will see a more advanced approach in section \ref{subsmap}). For a single binary trait with two states 0 and 1, there are two possible transitions: $0 \to 1$, and $1 \to 0$. For a pair of binary traits, there are four possible pairs of states, 00, 01, 10, and 11 and twelve possible transitions. \citet{ridley_explanation_1983} built a contingency table by counting the number of occurrences of each combination of traits at the end of branches where at least one trait had undergone a transition. As mutations on different branches of the tree can be considered independent, one can test independence using standard tests ($\chi^2$ or Fisher's exact test).

The method of \citet{ridley_explanation_1983} tests for covariation of traits, regardless the direction of mutation. It does not distinguish whether mutations in the first trait tend to cause mutations in the second. Such information can be inferred from branches where only one of the two traits undergoes a transition, noting the state of the other trait. \citet{maddison_method_1990} introduced a method where one of the two traits is treated as a conditional variable, and assessed whether the transitions at the other trait are distributed randomly according to the conditioned trait along the phylogeny. 

\citet{harvey_comparative_1991} extended the methods of \citet{ridley_explanation_1983} and \citet{maddison_method_1990} using a continuous time Markov model of trait evolution to compute the likelihood of the species distribution of states, given the underlying phylogeny. For binary traits, the generator of the Markov model is 
\begin{equation}
    Q = \begin{pmatrix}
    -q_{0\rightarrow 1} & q_{0\rightarrow 1} \\
    q_{1\rightarrow 0} & -q_{1\rightarrow 0}
    \end{pmatrix},
\end{equation}
where $q_{x \to y}$ denotes the rate of transition from state $x$ to state $y$.
This model is conceptually similar to the one introduced by \citet{felsenstein_evolutionary_1981} for modeling sequence evolution, from which it borrows the so-called pruning algorithm (dynamic programming) to efficiently integrate over all possible ancestral states.

One can then properly compute a likelihood for the mutations in the tree. The likelihood function can be maximised to estimate the transition matrix. Interestingly, \citet{harvey_comparative_1991} used this likelihood framework to compute the mean number of mutations on each branch of the phylogeny for each trait independently. A correlation coefficient can subsequently be computed to assess the independence between the evolution of the two traits.

This approach was further extended by \citet{pagel_detecting_1994}\label{sec:pagel1994}, who considered pairs of trait states in the Markov chain: 4 combined states and at most 12 transitions. More specifically, he used the following generator to model the correlated evolution of both traits:
\begin{equation}
    Q_\mathrm{pair} =
    \begin{pmatrix}
    -q_{00\to01} - q_{00\to10} & q_{00\to01} & q_{00\to10} & 0\\
    q_{01\to00} & -q_{01\to00} - q_{01\to11} & 0 & q_{01\to11}\\
    q_{10\to00} & 0 & -q_{10\to00} - q_{10\to11} & q_{10\to11}\\
    0 & q_{11\to01} & q_{11\to10} & -q_{11\to01} - q_{11\to10}
    \end{pmatrix}
\end{equation}
further assuming that the two traits cannot mutate simultaneously and thus reducing the transition matrix to 8 parameters. When the two traits are independent, the transition matrix can be even further simplify to 4 parameters, two transitions for each trait. \citep{pagel_detecting_1994} then suggested to compare the likelihood maximized over the 4 or 8 parameters (resp. $\mathcal{L}^{\max}_{4p}$ and $\mathcal{L}^{\max}_{8p}$). As both models are nested --the 4 parameter model is defined in a sub-space of the general one--, we can test whether the likelihood increase due to the extra parameters is significant. A model with more parameters always fits better, and the use of likelihood ratio tests is standard to compare nested models: under independence, the quantity $2(\ln{\mathcal{L}^{\max}_{8p}} - \ln{\mathcal{L}^{\max}_{4p}})$ converges to a $\chi
^2$ distribution with 4 degrees of freedom. The generality of this framework permits to test various hypotheses about the evolutionary process, including the order of mutations along the phylogeny. The relatively large number of parameters involved, however, restricts its application to relatively large data sets.


\subsection{Examples of correlated traits}

Applications of phylogenetic comparative analysis fall within two categories: (i) inferring the factors that drive the evolution of one or several traits, and (ii) assessing the correlated evolution between several traits. 
One of the most studied trait is the body mass index, which correlates with many other physiological and environmental variables \citep{cooper_body_2010}. It was mostly studied in mammals, for which large data sets are available, as well as a well-studied phylogeny. These studied revealed an important historical component \citep{gittleman_carnivore_1985, cheverud_quantitative_1985}, but also geographical and environmental factors such as temperature, altitude and species richness \citep{cooper_body_2010}. \citet{paradis_analysis_2002} used GEEs to reassess the relationship between species dispersal and population synchrony (defined as the correlation of population size variations between localities \citep{lande_spatial_1999}). It accounts for dispersal, habitat and the interaction of the two factors, together with a long term effect (time series) and the underlying phylogeny of the species. Their model confirms the existence of an habitat-specific effect of dispersal, but shows that the long term effect vanishes when phylogeny is properly accounted for.

Other studies used phylogenetic comparative analysis to assess correlated evolution between distinct traits. An example of such correlated traits is the average vs. sex dimorphism in species body mass. The ``Rensch rule'' \citep{Rensch1950DieAD} stipulates that sexual dimorphism increases with the average body mass in species where the male is the largest, while it decreases in species where the female is larger that the male \citep{fairbairn_allometry_1997}. \citet{abouheif_comparative_1997} used phylogenetic independent contrasts to put this rule at test with 21 animal taxa. They found that, besides one single exception, allometric relationships follow the pattern of Rensch rule.
\citet{lenormand_recombination_2005} looked at another sexual dimorphism, the difference of recombination rate between females and males (heterochiasmy). They fitted a GEE model to a plant data set and assessed the factors driving the evolution of heterochiasmy. There, the trait under study was the ratio of sex-specific average recombination rate, and the explanatory variables included, in addition to the phylogeny of the species, the presence of sex chromosomes and the selfing rate.  This model allowed the authors to test the Haldane-Huxley hypothesis, which stipulates that in species where recombination is suppressed in one sex, this sex is always the heterogametic one. The pattern, however, does not hold for so-called heterochiasmate species, where both sexes recombine but at a distinct rate. \citet{lenormand_recombination_2005} showed that the presence of sex chromosomes had no significant effect on the difference of recombination rate between sexes, but that the selfing rate did, as predicted by a model with differential selection during the haploid phase \citep{lenormand_evolution_2003}.

\subsection{Jointly modeling traits and sequences}

An interesting complementary approach tested for correlated evolution between transition rates of molecular evolution and several morphological traits \citep{lartillot_phylogenetic_2011}. Rates of molecular evolution, such as dN, the rate of amino-acid changes (non-synonymous) or dS, the rate of synonymous mutation rate, are modeled as continuous traits. Although one can only hardly measure rates, even at the leaf of the tree, the sequences at the leaves of the trees are informative on the mutation processes. The mutational rates are modeled explicitly, as in classical phylogeny, but their rates vary as a Brownian motion, whereas they are held constant in standard phylogeny. The ambition of such models is to find morphological traits, or life history traits, that strongly correlate with molecular rates variation. \citet{lartillot_phylogenetic_2011} found that body mass and longevity positively correlate with the intensity of selection, proxied by the ratio dN/dS. The same framework was also used to explore the interplay between biased gene conversion, selection and other traits \citep{lartillot_interaction_2013}. Finally, it was also used to explore how the variation of mutation rates and their correlation to traits could bias the inference of the phylogenetic tree itself \citep{lartillot_joint_2012}.

\section{Correlated evolution within genomes}
Genetic sequences can be viewed as traits with simple, well-defined discrete states (4 states for each site). Unlike morphological traits, the mechanisms of their evolution are well understood, with explicit models of transitions along a phylogeny. As a result, increasingly complex models of sequence evolution have been developed \citep{yang_computational_2006}. Correlated evolution of the molecular traits can thus be investigated using statistical methods that are similar to the ones described in section 1 for morphological traits. The current section presents different approaches that have been conducted, classified according to the selection level that they consider, from single nucleotides to complete genomes.   

\subsection{Within genes, between nucleotides}

One can trace back the study of intra-molecular coevolution to the research on RNA molecules. When \citet{holley_structure_1965} sequenced the first transfer RNA (tRNA) molecule, they proposed three possible structures based on the occurrence of stretches of Watson-Crick pairs, forming putative double helices. When a second tRNA sequences was obtained, the only common possible structure was one with a ``cloverleaf'' shape, which was then proved to be common to all tRNAs \citep{westhof_transfer_2012}. This conservation of tRNA structure opened the way to comparative sequence analysis as a way to extract information regarding evolutionary constraints and, ultimately, biological functionality -- as envisioned by Crick \citep{cobb_60_2017}.

The Watson-Crick base-pairing, that associates A-U and G-C, leads to an important evolutionary prediction: if two positions in a RNA molecule are interacting to form a Watson-Crick pair within a double helix, the paired  nucleotides with states A-U, U-A, G-C and C-G have a similar fitness and are thus inter-exchangeable. Conversely, any of the other pairs leads to a structural instability and, therefore, a lower fitness. Consequently, when comparing multiple homologous paired positions in different genes, we expect a high frequency of A-U, U-A, G-C and C-G, and a low frequency, if any, of the other combinations.
This results in a two-site fitness landscape where Watson-Crick pairs are ``peaks'' connected by two mutation events, one at each site, as illustrated in Figure \ref{figure:fl}D \citep{dutheil_base_2010}. A resulting property of this landscape is that the fitness effect of a mutation at one site (and consequently, its probability of fixation) depends on the state at the interacting site. 
Such epistatic interactions result in a scenario of coevolution by compensation: a deleterious mutation can be compensated by another mutation at an interacting site, and the evolution of the two positions become correlated. This reasoning served as a basis for many molecular coevolution studies. Identifying correlated evolution within a gene tells something about the structural and functional constraints of the encoded molecule. 

This principle was first applied to the prediction of RNA structures, given the clear expectation of coevolution at sites located within double-stranded helices (see \url{http://www.rna.icmb.utexas.edu/CAR/} for a historical perspective). The comparative analysis of RNA sequences starts with a multiple sequence alignment. The most simple approaches initially counted the number of Watson-Crick pair occurrences in all pairs of alignment columns, while subsequently developed approaches used more elaborated statistical methods. One of the earliest but still popular method uses the so-called mutual information (MI) measure, defined as the sum of the Shannon entropy of the two positions, minus the joint entropy of the pair \citep{chiu_inferring_1991}:
\begin{multline}\label{eqn:mi}
MI(x, y) = -\sum_i f(x_i) \log (f(x_i)) \\
-\sum_j f(y_j )\log (f(y_j))\\
+ \sum_i\sum_j f(x_iy_j) \log(f(x_iy_j))
\end{multline}
where $x$ and $y$ are two columns in the alignments. $f(x_i)$ and $f(y_j)$ denote the frequencies of each nucleotide $i$ at site $x$ and each nucleotide $j$ at site $y$, respectively. Besides, $f(x_i y_j)$ denotes the frequency of each pair of nucleotides $i$ and $j$ at the two sites. Given that $\sum_i f(x_i) = 1$ and $\sum_j f(y_j) = 1$, we have $\sum_i f(x_i y_j) = f(y_j)$ and $\sum_j f(x_i y_j) = f(x_i)$.
Equation \ref{eqn:mi} can then be factorized and reorganized in:
\begin{equation}
MI(x, y) = \sum_i\sum_j f(x_i y_i) \log\left( {\frac{f(x_i y_j)}{f(x_i)\times f(y_j)}} \right). 
\end{equation}
If the two positions $x$ and $y$ are evolving independently, then $f(x_i y_i) = f(x_i)\times f(y_i)$, and $MI(x,y) = 0$. A $MI$ value departing from 0 is therefore a signature of non-independent evolution.

A difficulty arises, however, when trying to evaluate the significance of $MI(x,y)$ as sequences partially shares history (again represented by a phylogeny). Consequently, $MI$ values will generally be positive for most pairs of sites \citep{dutheil_detecting_2012} and properly assessing the null distribution of $MI$ values requires accounting for the phylogeny. This can be achieved by correcting the $MI(x, y)$ value using information from the full sequence alignment (assuming coevolving pairs are a minority among all possible pairs) or by simulations (see \citet{dutheil_detecting_2012} for a review). 

While it is possible to properly account for the evolutionary history of the species when computing the null-distribution of the $MI$ statistics, this statistics is based on the sequence patterns only and is intrinsically agnostic of the underlying evolutionary paths. Several authors, therefore, designed models of coevolving positions, related to the model introduced by \citet{pagel_detecting_1994} (see section 1), restricted to pairs of sites to avoid an excessively large number of parameters. For nucleotides, the model has 16 states and at most 240 transitions (but as for the original model of Pagel, the number of transitions can be reduced by banning double mutations and by imposing constraints on the reversibility of transitions). \citet{tillier_neighbor_1995, tillier_high_1998} were the first to design a nucleotide model, which they applied to ribosomal sequences. A convenient aspect of this framework is that it is straightforward to derive the expected model when both sites evolve independently. This independence model can be used as a null-model and compared to the correlated model, using for example likelihood ratio tests. It requires, however, to fit a model for all pairs of sites, which can be relatively resource demanding. \citet{yeang_detecting_rna_2007} used a similar model, which they combined with a hidden Markov chain along the alignment in order to capture spatial dependencies, for instance within secondary structure motifs. Finally, \citet{dib_evolutionary_2014} used a general Pagel model, and employed a Bayesian inference procedure in order to jointly estimate model parameters, but also the coevolutionary ``profile'' of coevolving states among possible state pairs, as well as the underlying phylogeny \citep{meyer_simultaneous_2019}. Such approaches, combining sophisticated statistical inference and efficient computational implementations are very powerful tools to study coevolutionary processes within sequences.

\subsection{Within proteins, between amino-acids}

Characterizing correlated evolution within proteins is intrinsically more challenging, because of the larger number of states. While the 20 amino-acid lead to a putative 400 possible pairs of states for two coevolving positions, all these combinations may not differ in their fitness. Because the 20 amino-acids have redundant biochemical properties, some may be interchangeable at some positions in the protein without any notable fitness effect.

The year 1994 was particularly important for the study of protein coevolution. \citet{shindyalov_can_1994} introduced a new probabilistic method that ``mapped'' substitutions on the branches of a phylogenetic tree in order to assess how likely two positions are to undergo substitutions on the same branches of the tree, if they were evolving independently. \citet{shindyalov_can_1994, gobel_correlated_1994} then asked how the three-dimensional structure of proteins impacts the occurrence of coevolution; in particular, whether residues in contact in the tertiary structure tend to have a correlated evolution. \citet{neher_how_1994} developed a different statistical framework to address the same question. While his approach does not account for the underlying phylogeny, it proposes an attempt to account for the biochemical properties of proteins. By weighting each amino-acid by a given biochemical index, he assessed how correlated are biochemical properties between sites in a protein. 

\citet{pollock_coevolving_1999} used an approach similar to that employed by \citet{tillier_high_1998} to model the correlated evolution of RNA sites, but for proteins. In order to reduce the complexity of the model, which would contain a total of 400 states, they grouped several amino-acids according to their biochemical properties. If two categories are considered (for instance ``large'' and ``small'' residues), the model only has four states (large-large, large-small, small-large, small-small) to account for. It is similar to the general model introduced by \citet{pagel_detecting_1994}, yet the authors lowered the number of parameters (by making the model reversible).

\begin{figure}[ht!]
    \centering
    \includegraphics[width=\textwidth]{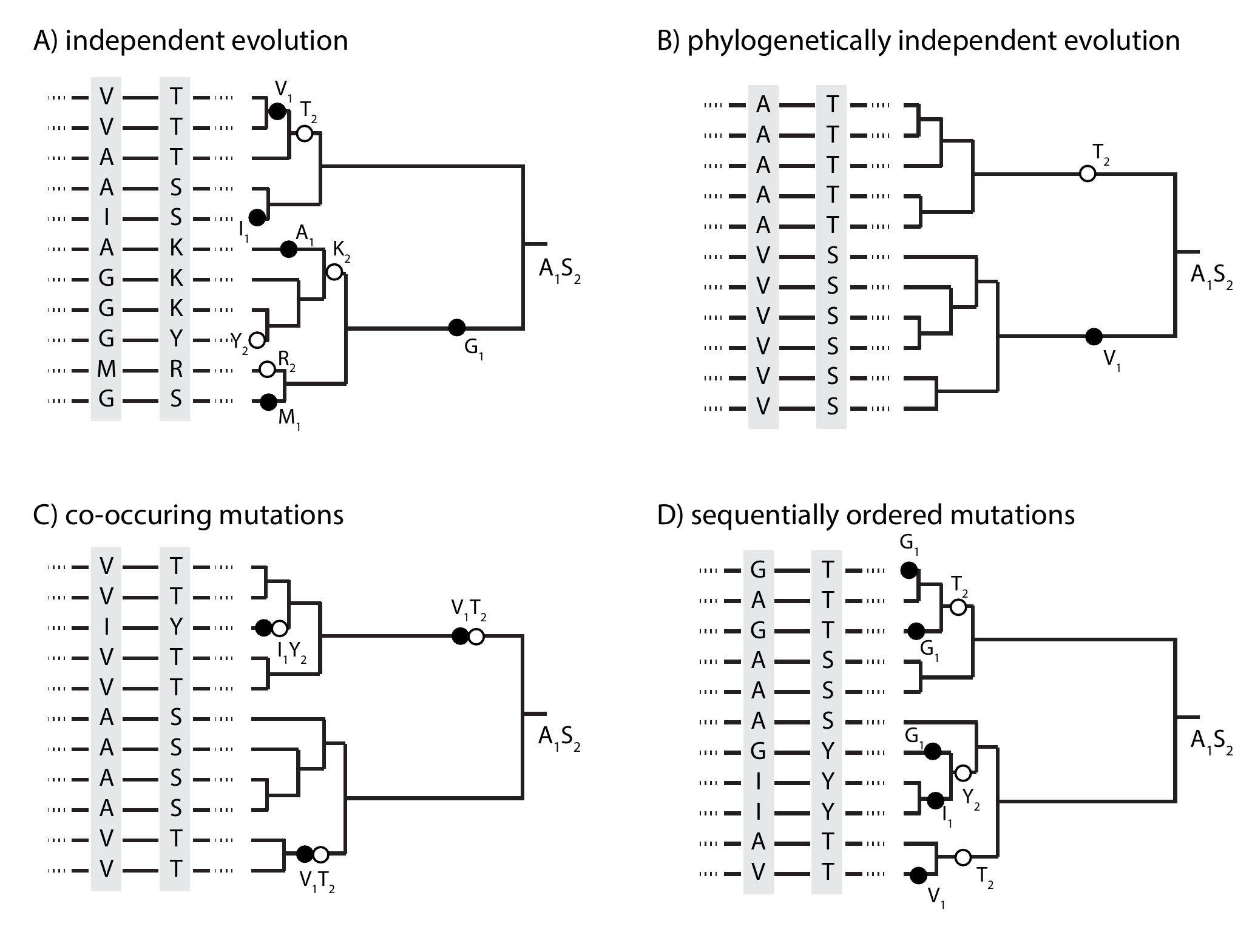}
    \caption{\small{Four possible outcomes of the comparison between two columns of an alignment. The grey area brings the focus to two columns of a larger alignment. The phylogenetic tree on the side illustrates the process of substitution mapping on the tree. Ancestral state is given at the root, mutations at the first site are displayed by black disks, while mutations at the second site by empty circles. A) For most pairs of sites, no correlated variation is observed neither in the alignment nor in the tree. B) Although the two pairs evolve independently, the phylogenetic inertia can mimic patterns of correlated variation in the alignment. The pattern would only be statistically significant for MI-like methods when phylogeny is ignored. C) A pattern of co-variation corresponding to co-substitutions in the tree, suggestive of a strong reciprocal interaction between both sites. D) A pattern of correlated evolution with a time lag between the first mutation and the following one, suggestive of an asymmetrical induction of mutations at site 2 on mutations at site 1.}}
    \label{fig:comap}
\end{figure}

\citet{tuffery_exploring_2000} introduced an intermediate approach. Building on \citet{shindyalov_can_1994}'s work, they first used a phylogeny to reconstruct the ancestral sequences at all inner nodes in the tree, and inferred on which branch and site substitutions occurred. They then evaluated which positions in the alignment tended to undergo substitutions on the same branches of the tree, introducing the term \emph{co-substitution} to denote branches with mutations at both sites. \citet{dutheil_model-based_2005} and \citep{dutheil_detecting_2007} later extended this method to account for the branch lengths of the tree and the uncertainty of the ancestral states, a procedure called \emph{substitution mapping}\label{subsmap}. They further introduced the possibility to include biochemical weights, following \citet{neher_how_1994}. A complementary development considers not only co-substitutions, but also sequentially ordered events (see Figure \ref{fig:comap}d), where substitutions at one site (the leading events) trigger substitutions at another site (the lagging events) \citep{kryazhimskiy_prevalence_2011, behdenna_testing_2016}. Once mutation events are placed on the tree, it is possible to compute analytically the probabilities under independence of any type of pairs (co-substitutions, ordered pairs, exclusive pairs, etc.) \citep{behdenna_testing_2016}. Such semi-parametric approaches are more efficient than full-model based approaches as they do not explicitly model site dependence. Yet the underlying models account for many aspects of the evolutionary process, and can be applied to both nucleic acids and proteins \citep{dutheil_model-based_2005}.

Parallel to the development of phylogenetically grounded methods, methods solely based on the alignment have recently shown spectacular development. A popular and successful approach is to fit a generalized Ising model (Potts model) into the alignment of $L$ amino-acid sequences, disregarding the phylogeny. Ising models consist of discrete variables that represent atomic "spins" that can be in one of two states (+1 or -1). The spins are arranged in a graph, usually a lattice, allowing each spin to interact with its neighbors (pairwise interactions). Generally, neighbors spins are only compatible when of the same sign. One can then compute the probability of the whole graph to be in a configuration using marginal probabilities of each spin and pairwise interactions between them. The Potts model allows for more than two states (not only +1/-1). These models, developed in statistical physics, take into account local abundances and pairwise interactions, but disregard higher order interactions. DCA (Direct Coupling Analysis) \citep{weigt_identification_2009, morcos_direct-coupling_2011} is a  successful implementation of a method based on this model. 

The Potts model assumes that each sequence of the alignment has a probability of being sampled that is given by a Boltzmann-like distribution. The higher the probability, the more abundant is the sequence in nature. Low probability sequences correspond to deleterious combinations of the amino-acids that one should not encounter in nature. A sequence $a_1 a_2 \cdots a_n$ has a probability $P(a_1 a_2 \cdots a_L) = \frac 1 Z  \exp(-\mathcal{H}_{a_1 a_2 \cdots a_L})$ of being observed. $Z$ is a normalisation constant, equal to the integral of $\exp(-\mathcal{H})$ over the whole sequence space. The ``Hamiltonian'' function $\mathcal{H}$ has both a local abundance component $h$ and a pairwise epistatic component $e$. More precisely, it is computed as:
$$\mathcal{H}_{a_1 a_2 \cdots a_L}= \sum_{i=1}^L h_i(a_i) + \sum_{i=1}^L \sum_{j=i+1}^L e_{ij}(a_i,a_j)  $$
The sums are taken over all positions of the alignment. In this formulation, the function $h$ takes $21 \times L$ input variables (gap is a 21st symbol) and returns a quantity related to the marginal probability of observing one of the 21 symbols at position $i$. The $e$ function takes $(21\times L)^2$ input variables and returns a quantity related to the symmetrical interaction between amino-acids (and gaps) for each pair of positions $i,j$, with $e_{ij}(a,b) = e_{ji}(b,a)$ and $e_{ii}(a,b)=0$.

One reason proposed to justify the Potts model for protein sequences is that the correlation observed between two columns in an alignment is not a good indicator of a direct functional or structural interactions. In the case of structure for instance, a correlation can be the result of a contact between two residues (a true signal of correlated evolution), but can also reflect indirect interactions: e.g. $a_i$ and $a_j$ are distant in the structure, but close to a third residue $a_k$, then resulting in indirect interactions. In the Potts model, the probability distribution only considers direct coupling effects. The hope is that those couplings reflect biologically relevant interactions. In this regard, only the $e$ function is important.

In the end, the difficulty comes when the functions $h$ and $e$ must be inferred from the data, a challenge named the \textit{inverse Potts problem}. Several techniques were proposed, from simple moment estimations to maximisation of the $\mathcal{H}$ function in its gigantic parameter space \citep{weigt_identification_2009, morcos_direct-coupling_2011, baldassi_fast_2014, ekeberg_improved_2013}. The method performs especially well, when a very large number of ``pseudo-independent" sequences are provided in the alignment. Indeed, it is noteworthy to mention that, to correct for the phylogenetic inertia, closely related sequences (by default with $>70\%$ of identity) are grouped into clusters and each member of the cluster has a weight inversely proportional to the cluster size. This also correct for bias sampling of the protein families in the database, as some clades are typically over-represented (e.g. vertebrates). DCA was originally developed to predict structural contacts between amino-acids \citep{weigt_identification_2009}, but as also used to find contacts in RNA \citep{de_leonardis_direct-coupling_2015} or predict the effect of mutations on a query sequence \citep{figliuzzi_coevolutionary_2016}.

While most methods typically exploit either evolutionary paths in phylogenies or counts in the alignment, others searched a compromise between the two, with the explicit goal to maximise the input information while keeping the computation time low. In CAPS \citep{fares_novel_2006}, the authors designed a score for two sites that uses all pairwise comparisons of the sequences; it thus takes into account the pairwise distances matrix between the sequences but does not reconstruct a tree. In the BIS method \citep{dib_protein_2012} and BIS2 \citep{champeimont_coevolution_2016}, subsequent residues that show strong patterns of co-abundance are first merged into blocks. All pairs of blocks are then tested for co-evolution by computing a metric of co-abundance in the alignment (an MI-like score). The metric is finally recursively re-weighted by the co-abundances in all subtrees, starting with the root node (the root node encompasses all sequences). Consequently, the topology of the tree (not the branch lengths) is incorporated in the metric. Finally more recently, persistence time in the phylogeny of co-conservations (not co-variations) for a pair of sites has been suggested to be a good sign of co-evolving positions \citep{laine_gemme_2019}.

Residues showing strong patterns of correlated evolution are typically in contact in the 3D structure of the folded protein \citep{pollock_coevolving_1999}. It is noteworthy to mention that coevolving residues detected by methods geared to direct coupling such as DCA-like are typically closer in space than methods solely based on correlation of abundances in the alignment, such as MI-like \citep{marks_protein_2011}. 
Conversely, it has been proposed to use the graph of co-evolving residues to guide the folding of peptides, a major challenge of structural biology. Recently developed folding methods, such as in the EVfold suite, use the predictions of DCA to greatly improve the folding accuracy \citep{marks_protein_2011, hopf_three-dimensional_2012}. 

The continuous development of innovative methods to study patterns of covariation in biological sequences has led to a better understanding of the relation between coevolution of sites and their functional linkage. Yet, several theoretical aspects of sequence coevolution remains poorly understood. What are the mechanisms by which co-substitutions occur? What is the role of compensatory mutations in sequence coevolution? What is the connection between coevolution and the rate of evolution at single sites? Answering these questions requires input from related fields of study, such as empirical fitness assessment \citep{visser_empirical_2014}, experimental evolution \citep{long_elucidating_2015}, as well as modeling \citep{talavera_covariation_2015}.

\subsection{Within genomes, between genes}

Cellular functions are the results of multiple protein interactions. Many proteins interact with more than one partner, resulting in a network of protein interactions. As proteins involved in the same network are likely to share some selective constraints, they are expected to coevolve, similarly to sites within individual proteins. Inferring the interaction partners of a protein is of great interest to elucidate the functional role of a candidate gene. Therefore, several studies aimed at inferring correlated evolution to predict protein-protein interactions \citep{de_juan_emerging_2013}.

A pioneering method, known as \textit{mirror-tree} detects genes with similar phylogenetic trees \citep{goh_co-evolution_2000, pazos_similarity_2001}. In its early implementation, the method simply looks at the correlation between distance matrices computed from two protein sequence alignments: because of common evolutionary constraints of interacting proteins, their rate of evolution will be correlated as they adapt simultaneously to the changing environment and, therefore, accumulate substitutions in the same branches of the tree. These assumptions are ``co-departures from a clockwise evolution'' \citep{galtier_coevolution_2007}. Yet, neutrally evolving proteins, which evolve under a molecular clock, will show a strong correlation of their genetic distances as they share a common history of speciation events. Accounting for this common phylogenetic context has been the focus of later developments \citep{de_juan_emerging_2013}.

With the advent of genome sequencing, exhaustive information regarding the proteome content of species became available. This allowed the reconstruction of \textit{phylogenetic profiles} (= \textit{phyletic patterns}), defined as the patterns of presence/absence of homologous genes in a set of species. \citet{barker_predicting_2005} applied the method of \citet{pagel_detecting_1994} (see Section \ref{sec:pagel1994}) to phyletic patterns from more than 10,000 protein pairs in order to detect linked genes. Using a set of known interacting proteins from \textit{Saccharomyces cerevisiae}, they found that functionally interacting pairs indeed show a signal of coevolution, and that using phylogeny-aware coevolution dectection methods significantly improves the prediction of functional interactions. The ``discrete coevolution'' model, however, requires several transition parameters to be estimated for each tested pair of proteins. \citet{barker_constrained_2007} later showed that fixing some of these parameters further improves the predictions. 

An alternative approach originates from the studies of the rates of genes gains and losses. These studies summarize the patterns of presence-absence of many gene families into a large matrix of 0 and 1, with species as rows and genes as columns, similar to a sequence alignment. Using the model introduced by \citet{harvey_comparative_1991} for single binary characters, it is possible to infer rates of gains and losses by maximum likelihood. In this approach, the number of parameters to estimate is considerably reduced, as they are shared by all genes, which resolves some of the estimation issues when fitting the same model on each gene independently. Yet, it also makes unrealistic assumptions, as some gene families are know to undergo higher rates of gene gains and losses than others. To accommodate this issue, \cite{cohen_likelihood_2008} proposed a mixture model allowing for these rate parameters to follow an \textit{a priori} distribution and, therefore, to vary between genes while introducing only a few additional parameters. Applying the substitution mapping approach introduced for mapping nucleotide and amino-acid substitutions \citep{dutheil_model-based_2005} to the presence-absence model, \citet{cohen_uncovering_2012} inferred events of co-gain and co-losses along the phylogeny to infer co-evolving genes. Using simulations, they show that this approach largely outperforms simpler methods based on the correlation of patterns only. The authors also showed that it gave slightly better performances than the \citet{barker_predicting_2005} model, with less than 1\% of the computational time, owing to the lower number of parameters to estimate \citep{cohen_copap_2013}.

Finally, it is worth mentioning that the physical interaction between proteins involves biochemical mechanisms acting at the amino-acid level. As a matter of fact, DCA was first developed to predict contact residues between two peptides \citep{weigt_identification_2009} and then improved to detect contact residues between two proteins \citep{szurmant_inter-residue_2018}. Conversely, DCA outcomes can be used to predict how two domains interact \citep{hopf_three-dimensional_2012}. However, finding co-evolving sites between pairs of proteins greatly increase the computational load, $n_1\times n_2$ comparisons have to be assessed between two genes encoding $n_1$ and $n_2$ amino-acids, respectively. Therefore, few methods were subsequently developed to apply a DCA-like strategy at the scale of complete genomes, such as superDCA \citep{puranen_superdca_2018}. Obviously, MI based methods can also be used to search for interacting partners of proteins \citep{bitbol_inferring_2018}, overcoming the computational cost of the inverse Potts problem. Recently, an efficient implementation of MI was used to scan interactions between all pairs of nucleotides in alignments of complete bacterial genomes \citep{pensar_genome-wide_2019}.

Finally, it worth mentioning that the phylogenic approach was also applied to this problem. \citet{yeang_detecting_2007} tackled this challenge by introducing a model of coevolution with only one additional parameter compared to the independent model. This parameter penalises simple substitutions while favoring cosubstitutions. Combining data filtering and using a computer grid, the authors could evaluate 0.1 trillion pairs of potentially interacting sites, and found that most coevolving pairs are also functionally interacting. A general issue with the detection of inter-genic correlated evolution is the assumption of a unique phylogeny shared by the two interacting genes. While phenomena like horizontal gene transfers are implicitly incorporated in models of gene gains and losses, gene duplication events are usually not, and at best, duplicated genes are discarded from further analyses. This problem remains largely unaddressed (but see \citet{yeang_identifying_2008}), and calls for a unification of methods to detect correlated evolution with methods of tree reconciliation.

\section{Genetics is also correlated evolution}

In this final section, we suggest that the whole history of gene mapping, since its very beginning, can be included in the large view of correlated evolution. Genetics can be considered as a specific case of correlated evolution where the first trait is discrete --presence/absence of a variant in the genome-- and the second is a phenotype of interest. Geneticists search for positive correlations between a phenotypic trait that is observed and a genetic trait that is unknown. In genetics, there is obviously a very strong support that we should interpret these correlations as causalities, where the presence of the allele in the genome ultimately causes the phenotype. Most methodologies of genetic mapping rely on exhibiting the joint occurrence (\textit{i.e.} a correlation) of one or several alleles with the phenotype. Our ambition is not to give a full and thorough review of the history and methods in genetics, but instead to suggest that ideas of correlated evolution may well have been growing with the rise of genetics in the early 20th century. 

\subsection{In individuals}

After the rediscovery of Mendel work in 1900 \citep{moore_rediscovery_2001}, started the rise of gene mapping, with research programs such as the one of TH Morgan group \citep{morgan_theory_2018}. The ambition was both to isolate the factors that segregated with phenotypes of drosophilas, and then to infer the linear genetic map of these factors. The phenotypes at stake were mostly natural mutants that were easily identifiable, such as mutants in eye-coloration or wing patterns. The experimental protocol consists mainly in two steps.

First testing whether a trait would segregate according to Mendel laws (each allele has a 0.5 chance of being transmitted to a progeny) by counting the number of descendants with or without the trait of interest in a controlled mating. It is noteworthy to mention that the existence of dominant and recessive alleles in diploïds, described by \citet{mendel_hybrid_1866}, add a modest complication for characterizing the transmissions. One must cross an individual with a dominant trait (A/-) with an individual without (-/-) to observe 50\% of the progeny harboring the phenotype. The observation that a trait would be transmitted with known and explicit rules demonstrates that a yet unknown genetic factor perfectly correlates with the presence of the trait. Interestingly the correlation between the allele and the trait is uniquely demonstrated by the observed segregating ratio.

The second part of Morgan's group work consists in controlled matings between males and females with two phenotypic traits. Using counts of co-occurrence between two traits in the progeny, they inferred the "linkage" between the corresponding genetic loci. A very simple estimation procedure of the linkage between two loci encoding two different dominant traits (say A and B) relies on counting in the progeny how many times a female\footnote{males do not recombine in drosophilas} with both traits in linkage (an AB phenotype, corresponding to an AB/-- genotype) mated with a male with no phenotype (genotype --/--) would produce progenies with 0, 1 or the 2 traits. When the original combination is kept, only descendants with 0 or 2 traits are observed. The mere observation of descendants with a single trait demonstrates the existence of new combinations of alleles that were named \textit{recombinations}. A first estimator of the recombination probability is the observed fraction of descendants with a single trait. Consequently, recombination rate, computed as such, never exceed 0.5: traits mapped to independent loci have a 0.5 chance of being co-transmitted. The general method relied on observing correlation between traits to infer correlation between loci.

By combining the pairwise measures of recombination rate, the group showed that the loci underlying the traits were arranged linearly in what they called linkage groups. The order and the distances between the loci can be computed using pairwise recombination rates by the transformation: $d=-50 \ln(1-2r)$, where $r$ is the recombination rate. These distances, named centiMorgan (cM), transform the $[0,0.5]$ range of recombination rate to proper $[0,\infty]$ distances. 

Although controlled matings were further done with almost all living organisms (including bacteria), it was not possible for humans, for obvious ethical reasons. Therefore, to map traits to loci in humans, researchers had to rely on other methods such as pedigree analysis.

\subsection{In pedigrees}

\begin{figure}[h]
	\centering
	\includegraphics[scale=0.45]{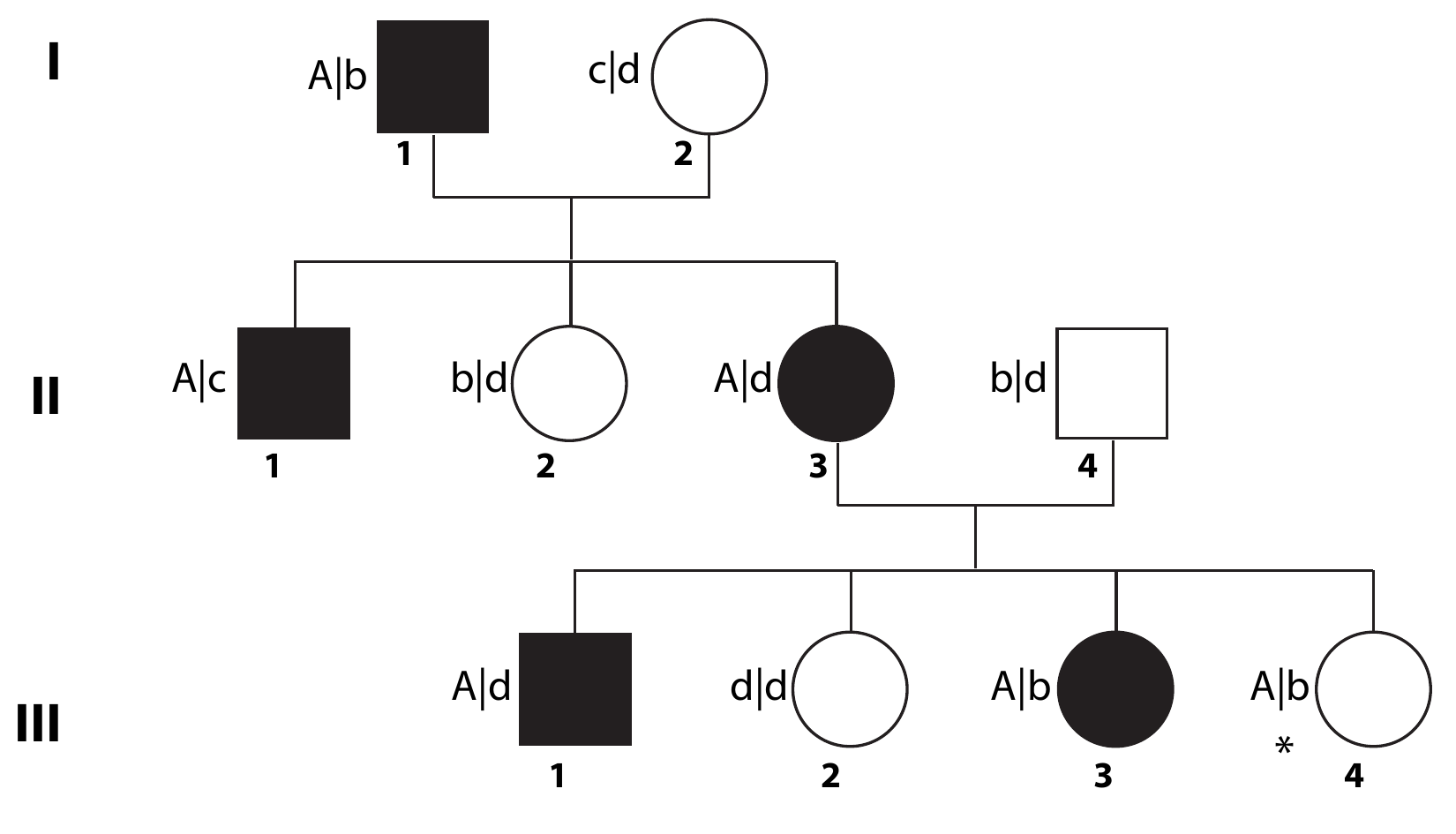} 
	\caption{\small{This pedigree reports the transmission of a phenotypic trait segregating through 3 generations, labeled I to III. Males (squares) and females (circles) are  colored in black when they show the phenotype. On the left of each individual, we report the observed genotype at a focal locus. Here, the suspicion is that the A allele at the locus co-segregates with the phenotype.}} 
	\label{pedigree}
\end{figure}

Pedigree analysis turned out to be a major tool in human genetics until the rise of association mapping in the 21st century. In a pedigree analysis, geneticists follow the segregation of a phenotypic trait, a pathology in medical  genetics, over two, three or ideally more generations. The trait is often binary: ill or not. Pedigree analysis is mostly suited to track dominant alleles. An example of a pedigree is presented in Figure \ref{pedigree}.

An implicit assumption of a pedigree analysis is that the phenotype is caused by an allele located at a single locus. By following the segregation of the trait, we infer the segregation of the causative allele. Using genotyping technologies, one can then genotype many loci throughout the genome for alleles that display a strong, possibly maximal, positive correlation with the phenotype. Geneticists leverage the correlated segregation between an observable phenotypic trait and a fully determined genotypic alleles to ``map'' a causative allele in a region of the genome. Until recently, only chosen markers distributed in genome were genotyped; consequently, the alleles co-segregating with the phenotype  were not causal, but only genetically linked with an unseen causal variant located nearby. Nowadays, as complete genome sequences are available, the causal allele is among the set of alleles that co-segregate with the trait.

The most popular metrics of co-segregation is the \textit{lod score} \citep{morton_sequential_1955}, a $\log$ of an odd-ratio. This odd ratio divides the likelihood of the observed co-segregation assuming linkage ($r<0.5$) to the likelihood of the observed co-segregation assuming independence ($r=0.5$):

$$S_r = \log_{10}\left( \frac{\mathcal{L}_{\mathbf{P}}({r<0.5})}{\mathcal{L}_{\mathbf{P}}(r=0.5) } \right)$$

To compute the likelihood, one follow the line of transmission of the allele with the phenotype. For example, in figure \ref{pedigree}, if one assumes that allele $A$ is linked to a causal allele located at distance $r$, the probability of observing the genotype of individual II.1, knowing that he has the phenotype, is $(1-r).\frac12$: $A$ did not recombined from the causal variant ($(1-r)$) and he received the allele $c$ from his mother ($\frac12$). The genotype of individual II.2 has probability $\frac14$. In individual III.4, there must have been a recombination between the allele $A$ and the causal variant; the probability is then $r.\frac12$. Overall, we have:
$$\mathcal{L}_{\mathbf{P}}(r) = \frac { (1-r)^3 r} {2^{10}} $$
More generally, the lod-score can be computed as
$$S_r = \log_{10} \left( {2^{NR+R} \times (1-r)^{NR} \times r^{R}} \right)$$
where $NR$ is the number of co-segregations of the focal allele with the phenotype (Non-Recombined) and $R$ the number of recombinations. The value of $r$ that maximizes the likelihood (and the lod-score) is the  natural estimate $\hat{r}_{ML}=R/(R+NR)$. By scanning the genome for strong associations, geneticists usually identify one region containing several markers with maximal lod-scores, named a haplotype. This region contains the unseen causal variant, which requires further sequencing and functional analyzes to be confirmed. It is usually recommended to reach a lod-score of 3 to claim a strong co-segregation \citep{nyholt_all_2000}: the odd of linkage is 1000 times times the odd of independence. 


\subsection{In the population}

In rare cases, multiple haplotypes correlate with the phenotype. However, when the trait is not digenic but polygenic, another corpus of methodology is more suited: the association mapping. Fundamental hypotheses of association mapping stems from quantitative genetics, where traits of interest are no more binary but continuous. The apparent continuity emerges from the combination of many loci of small effects, an assumption known as the \textit{infinitesimal model} \citep{barton_infinitesimal_2017}.

In association mapping, instead of a family, one uses the population phenotypic diversity to find independent genetic variants that contribute to the phenotype. For a binary phenotype, the assumption is that detected variants modulate the probability of having the phenotype, or that the pathology becomes visible only once a certain quantity of an invisible continuous trait is reached.








Association mapping at the scale of the genome is known as Genome Wide Association Study. In GWAS, geneticists measure the independent contribution of all variants segregating in a cohort to the phenotype of interest, assuming additive independent contribution for each locus. In a simple case, for a given locus $l$ hosting two alleles $A$ and $a$, a genotype $\mathcal{G}$ is 0, 1 or 2 copies of the $A$ allele. The statistical correlation between the number of $A$ in the genotype and the trait is computed either by a contingency table or by linear regression. 

{\bf Contingency table:} for a binary trait $\mathcal{T} \in \{0, 1\}$, we can construct a contingency table, where $n_{i,j}$ represent the number of individuals with genotype $\mathcal{G}=i$ and phenotype $\mathcal{T}=j$, as illustrated in Table \ref{table:contengency}.

\begin{table}[h]
\begin{center}
\begin{tabular}{|c|c|c|c|}
\hline
& $\mathcal{T}=0$ & $\mathcal{T}=1$\\
\hline
$\mathcal{G}=0$ & $n_{0,0}$ & $n_{0,1}$ \\
$\mathcal{G}=1$ & $n_{1,0}$ & $n_{1,1}$ \\
$\mathcal{G}=2$ & $n_{2,0}$ & $n_{2,1}$ \\
\hline
\end{tabular}
\end{center}
\caption{Contingency table for a binary trait and a bi-allelic locus.}
\label{table:contengency}
\end{table}%

Assuming co-dominance, the contingency table can be reduced to a 2x2 table by simply counting the number of alleles of type $A$ and $a$ associated to both traits: $n_{A,0}, n_{A,1}, n_{a,0}, n_{a,1}$: it is deduced from the general table with $n_{A,j} = 2n_{2,j}+n_{1,j}$ and $n_{a,j} = n_{1,j}+2n_{0,j}$. When one allele is dominant, the contingency table also shrinks to a 2x2 table by pooling the counts of genotype $\mathcal{G}=1$ with  the genotype $\mathcal{G}=2$ when $A$ is dominant ($n_{D,j}=n_{2,j}+n_{1,j};~n_{R,j}=n_{0,j}$). Association is then simply tested by a $\chi^2$ homogeneity test with 1 or 2 degrees of freedom depending on the table size. Larger contingency tables can be used to test for association for more than 2 alleles. Although the contingency table could, in theory, be extended to cases of non-binary phenotypes, it is in practice more convenient to re-express the problem in terms of linear regression.

{\bf Linear regression: }additive genetic effect can be estimated by the following linear model:
$$T_i = a_0 + a_1~G_i + \varepsilon_i $$
This model can be used estimate both (i) the significance and strength of the correlation of a given allele with the phenotype and (ii) $a_i$, the amount of additive effect of the allele to the trait. Techniques to estimate the $a_i$, such as least squares, are described in section 1. Although the basic version reported here assumes co-domminance, dominance can be explicitly addressed by setting the values of $G_i$ adequately. Linear models handle elegantly cases of continuous traits, but are not well adapted when more than two alleles exist at a given locus.

Both contingency tables and linear models can be used to estimate the association of alleles at a given locus to the trait. In GWAS, one assesses correlations with as many as possible polymorphic loci of the genome. The hope is to find a haplotype that is significantly associated to the trait. As there is a very large number of loci and thus of statistical tests, the power to detect a single association even in large samples is weak. Thus, GWASs need by construction very large sample sizes to potentially detect significant associations. In brief, basic GWAS is based on counting co-occurrences of alleles and traits in individuals, ignoring their shared history.

One inherent challenge of measuring independent contribution of the loci to the phenotype is the presence of epistasis between the loci. A phenotype can be the results of the synergistic interaction of multiple alleles (see Figure \ref{figure:fl}b). Correlations can hide correlations. Attempts to include epistasis in the detection of association include the use of Mutual Information (MI), generalized to 3-partners \citep{ignac_discovering_2014}, 2 genetic partners with symmetrical roles and 1 phenotypic trait that is the value to be predicted. In this model, both loci contribute together to the trait: they have an additive component and an interaction term, that is reminiscent of the DCA assumptions (see section 2). 

Although, erroneous associations could arise from multiple issues, one major annoyance is the stratification problem, that is the spurious association caused by population structuration \citep{lander_genetic_1994, pritchard_association_2000, price_principal_2006}. When a phenotype is mostly restricted to a subpopulation, all variants that characterize this subpopulation can be significantly associated with it. More generally, any correlation that is based on groups with and without a phenotype will suffer from the same bias, if there is some level of stratification. Therefore, like other correlated evolution methods based on counts of individuals, association mapping is biased by phylogenetic inertia, that can create non-relevant associations. Although there are several ways to handle this problem of stratification, one interesting method is to incorporate the genetic kinship between individuals in the model.

More specifically, individuals can be assigned to a genotype space, whose number of dimensions equals the number of genotyped loci. Their proximity in the genotype space relates to their kinship and their common ancestors \citep{mcvean_genealogical_2009}, as individuals with recent ancestors are more similar genetically. A convenient way to summarize the genotype space is to transform the original space using PCA \citep{novembre_interpreting_2008}. After the transformation, the first components of the new space are the ones explaining most of the variance, hence their name Principal Components, PC. Therefore, genetical proximity between individuals can be summarized (and often visualized) by only considering the first PCs that explain most of the genetic variance. One method \citep{price_principal_2006} correcting for stratification explicitly incorporates the positions of individuals, in the PCA space, within the linear model:

$$T_i = a_0 + a_1~G_i + \sum_j^n a_{j+1} \text{PC}_{i,j} + \varepsilon_i$$

where $\text{PC}_{i,j}$ is the value of the $j^{th}$ PC of individual $i$. Interestingly, PCA is also used to visually inspect the homogeneity between the individuals and eventually remove outliers from the analysis \citep{purcell_plink_2007}. It should be mentioned that other multidimensional transformations, such as MDS (multidimensional scaling), are sometimes preferred over PCA.

Although GWAS were first developed for humans \citep{ozaki_functional_2002}, they rapidly were applied to other organisms such as plants \citep{atwell_genome-wide_2010}, cattle \citep{mai_genome-wide_2010} or even, more recently, bacteria \citep{chen_advent_2015}. Interestingly, in bacteria the impact of stratification is even stronger as recombination in bacterial genomes differs strongly from the meiotic recombination in Eukaryotes. In bacteria, recombination is assimilated to gene conversion, as it replaces only locally a piece of the genome, without affecting its backbone. Thus the kinship of bacterial genomes within species can be described by a tree, especially for species with a low level of recombination (the tree paradigm suffers major drawback when recombination is high or when divergence between lineages is large). Therefore, GWAS methods in bacteria were redesigned to correct for the genealogy of genomes, using methods almost identical to substitution mapping reported in section 2. However, there, all sites of the core genome are mapped on the tree together with the trait of interest. And, instead at looking at correlated evolution between all pairs of sites, the correlation is computed between each site and the trait \citep{collins_phylogenetic_2018}. The observation that treeWAS, the recent GWAS version that is tree-aware, uses standard metric of correlated evolution that are precisely the same than the ones used to detect correlated evolution among sites in a genome, highlights again the theoretical and methodological proximity between classical genetics and correlated evolution.

\section{Conclusion}
In this chapter, we have reviewed several models and methods used to characterize correlated evolution. We have divided the chapter in 3 sections, covering coevolution within morphological traits, coevolution within genetic traits and finally coevolution between genetic alleles and phenotypic traits. As we have seen, most methods we have described are highly similar in these three fields of applications. Even more, several methods can be directly applied to any types of traits, provided they are re-coded in the same abstraction. This unity in the methodologies fosters the common underlying framework of all these applications. We hope that this review will help readers to connect fields that could at first sight be considered as unrelated. Models and methods evolve, similar to living organisms. Consequently, they also share common ancestries, explaining their relatedness.

We have described two main schools of modelling correlated evolution that were developed in parallel. The first one simply considers sampled species, individuals or sequences, disregarding their shared history, and compute various metrics of correlation among the traits. A second one explicitly uses the shared history to compute correlation among evolutionary or segregation paths of the traits. We shall emphasize that, without any doubt, both frameworks have been successfully applied to many different questions, demonstrating their respective inner strengths. We would also like to stress that there may be good reasons to use one or the other. Practically, it can be hard to assess with confidence a phylogeny. Indeed, many problems can arise like: a) individuals within species have not a single genealogy for their their genome, making difficult to infer the tree sequence, b) inference can be entailed with large errors, especially for very divergent species, c) computation time can be too high in very large samples to infer a tree. Thus, in many cases, simply counting co-occurrences without looking at the phylogeny is an excellent choice. Conversely, when the genealogy can be assessed with confidence, computing correlations between evolutionary paths is powerful, as it follows the evolutionary processes rather solely their final outcomes. Typically on datasets of small size, the phylogenetic methods often perform better.

Interestingly, we have also seen that despite the change of traits one is willing to study (morphological, genetics, etc.), the fundamental principles and models of the methods remained invariant. However, as the data grow in number and in quality, the fine implementation of the methods has been largely improved. 

We have show that the method of correlating traits in biology to deduce some causality using prior knowledge between a predictor trait (e.g. a gene) and a induced one (e.g. a phenotype) is an old practice; this, despite the rampant ambiguity between correlation and causation. It is tempting to remind that formal statistical correlations were popularized under the impulse of biometricians \citep{galton_regression_1886, pearson_vii_1895}, suggesting that correlated evolution in biology has more ancient roots than the rise of genetics.

\section{Ackowledgments}
GA is supported by the INCEPTION program of the Pasteur Institute.

\bibliography{references}

\begin{thebibliography}{104}
\providecommand{\natexlab}[1]{#1}
\providecommand{\url}[1]{\texttt{#1}}
\expandafter\ifx\csname urlstyle\endcsname\relax
  \providecommand{\doi}[1]{doi: #1}\else
  \providecommand{\doi}{doi: \begingroup \urlstyle{rm}\Url}\fi

\bibitem[Abouheif and Fairbairn(1997)]{abouheif_comparative_1997}
E.~Abouheif and D.~J. Fairbairn.
\newblock A {Comparative} {Analysis} of {Allometry} for {Sexual} {Size}
  {Dimorphism}: {Assessing} {Rensch}'s {Rule}.
\newblock \emph{The American Naturalist}, 149\penalty0 (3):\penalty0 540--562,
  1997.
\newblock ISSN 0003-0147.
\newblock URL \url{https://www.jstor.org/stable/2463382}.
\newblock Publisher: [University of Chicago Press, American Society of
  Naturalists].

\bibitem[Achaz et~al.(2014)Achaz, Rodriguez-Verdugo, Gaut, and
  Tenaillon]{achaz_reproducibility_2014}
G.~Achaz, A.~Rodriguez-Verdugo, B.~S. Gaut, and O.~Tenaillon.
\newblock The {Reproducibility} of {Adaptation} in the {Light} of
  {Experimental} {Evolution} with {Whole} {Genome} {Sequencing}.
\newblock In C.~R. Landry and N.~Aubin-Horth, editors, \emph{Ecological
  {Genomics}: {Ecology} and the {Evolution} of {Genes} and {Genomes}}, Advances
  in {Experimental} {Medicine} and {Biology}, pages 211--231. Springer
  Netherlands, Dordrecht, 2014.
\newblock ISBN 978-94-007-7347-9.
\newblock \doi{10.1007/978-94-007-7347-9_11}.
\newblock URL \url{https://doi.org/10.1007/978-94-007-7347-9_11}.

\bibitem[Atwell et~al.(2010)Atwell, Huang, Vilhjálmsson, Willems, Horton, Li,
  Meng, Platt, Tarone, Hu, Jiang, Muliyati, Zhang, Amer, Baxter, Brachi, Chory,
  Dean, Debieu, de~Meaux, Ecker, Faure, Kniskern, Jones, Michael, Nemri, Roux,
  Salt, Tang, Todesco, Traw, Weigel, Marjoram, Borevitz, Bergelson, and
  Nordborg]{atwell_genome-wide_2010}
S.~Atwell, Y.~S. Huang, B.~J. Vilhjálmsson, G.~Willems, M.~Horton, Y.~Li,
  D.~Meng, A.~Platt, A.~M. Tarone, T.~T. Hu, R.~Jiang, N.~W. Muliyati,
  X.~Zhang, M.~A. Amer, I.~Baxter, B.~Brachi, J.~Chory, C.~Dean, M.~Debieu,
  J.~de~Meaux, J.~R. Ecker, N.~Faure, J.~M. Kniskern, J.~D.~G. Jones,
  T.~Michael, A.~Nemri, F.~Roux, D.~E. Salt, C.~Tang, M.~Todesco, M.~B. Traw,
  D.~Weigel, P.~Marjoram, J.~O. Borevitz, J.~Bergelson, and M.~Nordborg.
\newblock Genome-wide association study of 107 phenotypes in {Arabidopsis}
  thaliana inbred lines.
\newblock \emph{Nature}, 465\penalty0 (7298):\penalty0 627--631, June 2010.
\newblock ISSN 1476-4687.
\newblock \doi{10.1038/nature08800}.
\newblock URL \url{https://www.nature.com/articles/nature08800}.
\newblock Number: 7298 Publisher: Nature Publishing Group.

\bibitem[Baldassi et~al.(2014)Baldassi, Zamparo, Feinauer, Procaccini,
  Zecchina, Weigt, and Pagnani]{baldassi_fast_2014}
C.~Baldassi, M.~Zamparo, C.~Feinauer, A.~Procaccini, R.~Zecchina, M.~Weigt, and
  A.~Pagnani.
\newblock Fast and accurate multivariate {Gaussian} modeling of protein
  families: predicting residue contacts and protein-interaction partners.
\newblock \emph{PLoS ONE}, 9\penalty0 (3):\penalty0 e92721, 2014.
\newblock ISSN 1932-6203.
\newblock \doi{10.1371/journal.pone.0092721}.

\bibitem[Barker and Pagel(2005)]{barker_predicting_2005}
D.~Barker and M.~Pagel.
\newblock Predicting functional gene links from phylogenetic-statistical
  analyses of whole genomes.
\newblock \emph{PLoS Comput. Biol.}, 1\penalty0 (1):\penalty0 e3, June 2005.
\newblock ISSN 1553-734X.
\newblock \doi{10.1371/journal.pcbi.0010003}.

\bibitem[Barker et~al.(2007)Barker, Meade, and Pagel]{barker_constrained_2007}
D.~Barker, A.~Meade, and M.~Pagel.
\newblock Constrained models of evolution lead to improved prediction of
  functional linkage from correlated gain and loss of genes.
\newblock \emph{Bioinformatics}, 23\penalty0 (1):\penalty0 14--20, Jan. 2007.
\newblock ISSN 1367-4811.
\newblock \doi{10.1093/bioinformatics/btl558}.

\bibitem[Barton et~al.(2017)Barton, Etheridge, and
  Véber]{barton_infinitesimal_2017}
N.~H. Barton, A.~M. Etheridge, and A.~Véber.
\newblock The infinitesimal model: {Definition}, derivation, and implications.
\newblock \emph{Theoretical Population Biology}, 118:\penalty0 50--73, Dec.
  2017.
\newblock ISSN 0040-5809.
\newblock \doi{10.1016/j.tpb.2017.06.001}.
\newblock URL
  \url{http://www.sciencedirect.com/science/article/pii/S0040580917300886}.

\bibitem[Behdenna et~al.(2016)Behdenna, Pothier, Abby, Lambert, and
  Achaz]{behdenna_testing_2016}
A.~Behdenna, J.~Pothier, S.~S. Abby, A.~Lambert, and G.~Achaz.
\newblock Testing for {Independence} between {Evolutionary} {Processes}.
\newblock \emph{Syst. Biol.}, 65\penalty0 (5):\penalty0 812--823, Sept. 2016.
\newblock ISSN 1076-836X.
\newblock \doi{10.1093/sysbio/syw004}.

\bibitem[Bitbol(2018)]{bitbol_inferring_2018}
A.-F. Bitbol.
\newblock Inferring interaction partners from protein sequences using mutual
  information.
\newblock \emph{PLOS Computational Biology}, 14\penalty0 (11):\penalty0
  e1006401, Nov. 2018.
\newblock ISSN 1553-7358.
\newblock \doi{10.1371/journal.pcbi.1006401}.
\newblock URL
  \url{https://journals.plos.org/ploscompbiol/article?id=10.1371/journal.pcbi.1006401}.
\newblock Publisher: Public Library of Science.

\bibitem[Blomberg et~al.(2003)Blomberg, Garland, and
  Ives]{blomberg_testing_2003}
S.~P. Blomberg, T.~Garland, and A.~R. Ives.
\newblock Testing for phylogenetic signal in comparative data: behavioral
  traits are more labile.
\newblock \emph{Evolution}, 57\penalty0 (4):\penalty0 717--745, Apr. 2003.
\newblock ISSN 0014-3820.
\newblock \doi{10.1111/j.0014-3820.2003.tb00285.x}.

\bibitem[Champeimont et~al.(2016)Champeimont, Laine, Hu, Penin, and
  Carbone]{champeimont_coevolution_2016}
R.~Champeimont, E.~Laine, S.-W. Hu, F.~Penin, and A.~Carbone.
\newblock Coevolution analysis of {Hepatitis} {C} virus genome to identify the
  structural and functional dependency network of viral proteins.
\newblock \emph{Sci Rep}, 6\penalty0 (1):\penalty0 1--20, May 2016.
\newblock ISSN 2045-2322.
\newblock \doi{10.1038/srep26401}.
\newblock URL \url{https://www.nature.com/articles/srep26401}.
\newblock Number: 1 Publisher: Nature Publishing Group.

\bibitem[Chen and Shapiro(2015)]{chen_advent_2015}
P.~E. Chen and B.~J. Shapiro.
\newblock The advent of genome-wide association studies for bacteria.
\newblock \emph{Current Opinion in Microbiology}, 25:\penalty0 17--24, June
  2015.
\newblock ISSN 1369-5274.
\newblock \doi{10.1016/j.mib.2015.03.002}.
\newblock URL
  \url{http://www.sciencedirect.com/science/article/pii/S1369527415000375}.

\bibitem[Cheverud et~al.(1985)Cheverud, Dow, and
  Leutenegger]{cheverud_quantitative_1985}
J.~M. Cheverud, M.~M. Dow, and W.~Leutenegger.
\newblock The quantitative assessment of phylogenetic constraints in
  comparative analyses: sexual dimorphism in body weight among primates.
\newblock \emph{Evolution}, 39\penalty0 (6):\penalty0 1335--1351, Nov. 1985.
\newblock ISSN 1558-5646.
\newblock \doi{10.1111/j.1558-5646.1985.tb05699.x}.

\bibitem[Chiu and Kolodziejczak(1991)]{chiu_inferring_1991}
D.~K. Chiu and T.~Kolodziejczak.
\newblock Inferring consensus structure from nucleic acid sequences.
\newblock \emph{Comput. Appl. Biosci.}, 7\penalty0 (3):\penalty0 347--352, July
  1991.
\newblock ISSN 0266-7061.
\newblock \doi{10.1093/bioinformatics/7.3.347}.

\bibitem[Cobb(2017)]{cobb_60_2017}
M.~Cobb.
\newblock 60 years ago, {Francis} {Crick} changed the logic of biology.
\newblock \emph{PLoS Biol.}, 15\penalty0 (9):\penalty0 e2003243, Sept. 2017.
\newblock ISSN 1545-7885.
\newblock \doi{10.1371/journal.pbio.2003243}.

\bibitem[Cohen et~al.(2008)Cohen, Rubinstein, Stern, Gophna, and
  Pupko]{cohen_likelihood_2008}
O.~Cohen, N.~D. Rubinstein, A.~Stern, U.~Gophna, and T.~Pupko.
\newblock A likelihood framework to analyse phyletic patterns.
\newblock \emph{Philos. Trans. R. Soc. Lond., B, Biol. Sci.}, 363\penalty0
  (1512):\penalty0 3903--3911, Dec. 2008.
\newblock ISSN 1471-2970.
\newblock \doi{10.1098/rstb.2008.0177}.

\bibitem[Cohen et~al.(2012)Cohen, Ashkenazy, Burstein, and
  Pupko]{cohen_uncovering_2012}
O.~Cohen, H.~Ashkenazy, D.~Burstein, and T.~Pupko.
\newblock Uncovering the co-evolutionary network among prokaryotic genes.
\newblock \emph{Bioinformatics}, 28\penalty0 (18):\penalty0 i389--i394, Sept.
  2012.
\newblock ISSN 1367-4811.
\newblock \doi{10.1093/bioinformatics/bts396}.

\bibitem[Cohen et~al.(2013)Cohen, Ashkenazy, Levy~Karin, Burstein, and
  Pupko]{cohen_copap_2013}
O.~Cohen, H.~Ashkenazy, E.~Levy~Karin, D.~Burstein, and T.~Pupko.
\newblock {CoPAP}: {Coevolution} of presence-absence patterns.
\newblock \emph{Nucleic Acids Res.}, 41\penalty0 (Web Server issue):\penalty0
  W232--237, July 2013.
\newblock ISSN 1362-4962.
\newblock \doi{10.1093/nar/gkt471}.

\bibitem[Collins and Didelot(2018)]{collins_phylogenetic_2018}
C.~Collins and X.~Didelot.
\newblock A phylogenetic method to perform genome-wide association studies in
  microbes that accounts for population structure and recombination.
\newblock \emph{PLOS Computational Biology}, 14\penalty0 (2):\penalty0
  e1005958, Feb. 2018.
\newblock ISSN 1553-7358.
\newblock \doi{10.1371/journal.pcbi.1005958}.
\newblock URL
  \url{https://journals.plos.org/ploscompbiol/article?id=10.1371/journal.pcbi.1005958}.
\newblock Publisher: Public Library of Science.

\bibitem[Cooper and Purvis(2010)]{cooper_body_2010}
N.~Cooper and A.~Purvis.
\newblock Body size evolution in mammals: complexity in tempo and mode.
\newblock \emph{Am. Nat.}, 175\penalty0 (6):\penalty0 727--738, June 2010.
\newblock ISSN 1537-5323.
\newblock \doi{10.1086/652466}.

\bibitem[de~Juan et~al.(2013)de~Juan, Pazos, and
  Valencia]{de_juan_emerging_2013}
D.~de~Juan, F.~Pazos, and A.~Valencia.
\newblock Emerging methods in protein co-evolution.
\newblock \emph{Nat. Rev. Genet.}, 14\penalty0 (4):\penalty0 249--261, Apr.
  2013.
\newblock ISSN 1471-0064.
\newblock \doi{10.1038/nrg3414}.

\bibitem[De~Leonardis et~al.(2015)De~Leonardis, Lutz, Ratz, Cocco, Monasson,
  Schug, and Weigt]{de_leonardis_direct-coupling_2015}
E.~De~Leonardis, B.~Lutz, S.~Ratz, S.~Cocco, R.~Monasson, A.~Schug, and
  M.~Weigt.
\newblock Direct-{Coupling} {Analysis} of nucleotide coevolution facilitates
  {RNA} secondary and tertiary structure prediction.
\newblock \emph{Nucleic Acids Res.}, 43\penalty0 (21):\penalty0 10444--10455,
  Dec. 2015.
\newblock ISSN 1362-4962.
\newblock \doi{10.1093/nar/gkv932}.

\bibitem[Dib and Carbone(2012)]{dib_protein_2012}
L.~Dib and A.~Carbone.
\newblock Protein {Fragments}: {Functional} and {Structural} {Roles} of {Their}
  {Coevolution} {Networks}.
\newblock \emph{PLOS ONE}, 7\penalty0 (11):\penalty0 e48124, Nov. 2012.
\newblock ISSN 1932-6203.
\newblock \doi{10.1371/journal.pone.0048124}.
\newblock URL
  \url{https://journals.plos.org/plosone/article?id=10.1371/journal.pone.0048124}.
\newblock Publisher: Public Library of Science.

\bibitem[Dib et~al.(2014)Dib, Silvestro, and Salamin]{dib_evolutionary_2014}
L.~Dib, D.~Silvestro, and N.~Salamin.
\newblock Evolutionary footprint of coevolving positions in genes.
\newblock \emph{Bioinformatics}, 30\penalty0 (9):\penalty0 1241--1249, May
  2014.
\newblock ISSN 1367-4811.
\newblock \doi{10.1093/bioinformatics/btu012}.

\bibitem[Dutheil and Galtier(2007)]{dutheil_detecting_2007}
J.~Dutheil and N.~Galtier.
\newblock Detecting groups of coevolving positions in a molecule: a clustering
  approach.
\newblock \emph{BMC Evol. Biol.}, 7:\penalty0 242, 2007.
\newblock ISSN 1471-2148.
\newblock \doi{10.1186/1471-2148-7-242}.

\bibitem[Dutheil et~al.(2005)Dutheil, Pupko, Jean-Marie, and
  Galtier]{dutheil_model-based_2005}
J.~Dutheil, T.~Pupko, A.~Jean-Marie, and N.~Galtier.
\newblock A model-based approach for detecting coevolving positions in a
  molecule.
\newblock \emph{Mol. Biol. Evol.}, 22\penalty0 (9):\penalty0 1919--1928, Sept.
  2005.
\newblock ISSN 0737-4038.
\newblock \doi{10.1093/molbev/msi183}.

\bibitem[Dutheil(2012)]{dutheil_detecting_2012}
J.~Y. Dutheil.
\newblock Detecting coevolving positions in a molecule: why and how to account
  for phylogeny.
\newblock \emph{Brief. Bioinformatics}, 13\penalty0 (2):\penalty0 228--243,
  Mar. 2012.
\newblock ISSN 1477-4054.
\newblock \doi{10.1093/bib/bbr048}.

\bibitem[Dutheil et~al.(2010)Dutheil, Jossinet, and Westhof]{dutheil_base_2010}
J.~Y. Dutheil, F.~Jossinet, and E.~Westhof.
\newblock Base pairing constraints drive structural epistasis in ribosomal
  {RNA} sequences.
\newblock \emph{Mol. Biol. Evol.}, 27\penalty0 (8):\penalty0 1868--1876, Aug.
  2010.
\newblock ISSN 1537-1719.
\newblock \doi{10.1093/molbev/msq069}.

\bibitem[Ekeberg et~al.(2013)Ekeberg, Lövkvist, Lan, Weigt, and
  Aurell]{ekeberg_improved_2013}
M.~Ekeberg, C.~Lövkvist, Y.~Lan, M.~Weigt, and E.~Aurell.
\newblock Improved contact prediction in proteins: using pseudolikelihoods to
  infer {Potts} models.
\newblock \emph{Phys Rev E Stat Nonlin Soft Matter Phys}, 87\penalty0
  (1):\penalty0 012707, Jan. 2013.
\newblock ISSN 1550-2376.
\newblock \doi{10.1103/PhysRevE.87.012707}.

\bibitem[Fairbairn(1997)]{fairbairn_allometry_1997}
D.~J. Fairbairn.
\newblock Allometry for {Sexual} {Size} {Dimorphism}: {Pattern} and {Process}
  in the {Coevolution} of {Body} {Size} in {Males} and {Females}.
\newblock \emph{Annu. Rev. Ecol. Syst.}, 28\penalty0 (1):\penalty0 659--687,
  Nov. 1997.
\newblock ISSN 0066-4162.
\newblock \doi{10.1146/annurev.ecolsys.28.1.659}.
\newblock URL
  \url{https://www.annualreviews.org/doi/10.1146/annurev.ecolsys.28.1.659}.
\newblock Publisher: Annual Reviews.

\bibitem[Fares and Travers(2006)]{fares_novel_2006}
M.~A. Fares and S.~A.~A. Travers.
\newblock A novel method for detecting intramolecular coevolution: adding a
  further dimension to selective constraints analyses.
\newblock \emph{Genetics}, 173\penalty0 (1):\penalty0 9--23, May 2006.
\newblock ISSN 0016-6731.
\newblock \doi{10.1534/genetics.105.053249}.

\bibitem[Felsenstein(1973)]{felsenstein_maximum-likelihood_1973}
J.~Felsenstein.
\newblock Maximum-likelihood estimation of evolutionary trees from continuous
  characters.
\newblock \emph{Am J Hum Genet}, 25\penalty0 (5):\penalty0 471--492, Sept.
  1973.
\newblock ISSN 0002-9297.
\newblock URL \url{https://www.ncbi.nlm.nih.gov/pmc/articles/PMC1762641/}.

\bibitem[Felsenstein(1981)]{felsenstein_evolutionary_1981}
J.~Felsenstein.
\newblock Evolutionary trees from {DNA} sequences: a maximum likelihood
  approach.
\newblock \emph{J. Mol. Evol.}, 17\penalty0 (6):\penalty0 368--376, 1981.
\newblock ISSN 0022-2844.

\bibitem[Felsenstein(1985)]{felsenstein_phylogenies_1985}
J.~Felsenstein.
\newblock Phylogenies and the {Comparative} {Method}.
\newblock \emph{The American Naturalist}, 125\penalty0 (1):\penalty0 1--15,
  1985.
\newblock ISSN 0003-0147.
\newblock URL \url{https://www.jstor.org/stable/2461605}.
\newblock Publisher: [University of Chicago Press, American Society of
  Naturalists].

\bibitem[Felsenstein(2003)]{felsenstein_inferring_2003}
J.~Felsenstein.
\newblock \emph{Inferring {Phylogenies}}.
\newblock Sinauer Associates, 2 edition, Sept. 2003.
\newblock ISBN 0-87893-177-5.

\bibitem[Figliuzzi et~al.(2016)Figliuzzi, Jacquier, Schug, Tenaillon, and
  Weigt]{figliuzzi_coevolutionary_2016}
M.~Figliuzzi, H.~Jacquier, A.~Schug, O.~Tenaillon, and M.~Weigt.
\newblock Coevolutionary {Landscape} {Inference} and the {Context}-{Dependence}
  of {Mutations} in {Beta}-{Lactamase} {TEM}-1.
\newblock \emph{Mol. Biol. Evol.}, 33\penalty0 (1):\penalty0 268--280, Jan.
  2016.
\newblock ISSN 1537-1719.
\newblock \doi{10.1093/molbev/msv211}.

\bibitem[Galtier and Dutheil(2007)]{galtier_coevolution_2007}
N.~Galtier and J.~Dutheil.
\newblock Coevolution within and between genes.
\newblock \emph{Genome Dyn}, 3:\penalty0 1--12, 2007.
\newblock ISSN 1660-9263.
\newblock \doi{10.1159/000107599}.

\bibitem[Galton(1886)]{galton_regression_1886}
F.~Galton.
\newblock Regression {Towards} {Mediocrity} in {Hereditary} {Stature}.
\newblock Jan. 1886.
\newblock \doi{10.2307/2841583}.
\newblock URL \url{https://zenodo.org/record/1449548#.XsfkzJrRaqA}.

\bibitem[Gittleman(1985)]{gittleman_carnivore_1985}
J.~L. Gittleman.
\newblock Carnivore body size: {Ecological} and taxonomic correlates.
\newblock \emph{Oecologia}, 67\penalty0 (4):\penalty0 540--554, Dec. 1985.
\newblock ISSN 1432-1939.
\newblock \doi{10.1007/BF00790026}.

\bibitem[Goh et~al.(2000)Goh, Bogan, Joachimiak, Walther, and
  Cohen]{goh_co-evolution_2000}
C.~S. Goh, A.~A. Bogan, M.~Joachimiak, D.~Walther, and F.~E. Cohen.
\newblock Co-evolution of proteins with their interaction partners.
\newblock \emph{J. Mol. Biol.}, 299\penalty0 (2):\penalty0 283--293, June 2000.
\newblock ISSN 0022-2836.
\newblock \doi{10.1006/jmbi.2000.3732}.

\bibitem[Grafen(1989)]{grafen_phylogenetic_1989}
A.~Grafen.
\newblock The phylogenetic regression.
\newblock \emph{Philos. Trans. R. Soc. Lond., B, Biol. Sci.}, 326\penalty0
  (1233):\penalty0 119--157, Dec. 1989.
\newblock ISSN 0962-8436.
\newblock \doi{10.1098/rstb.1989.0106}.

\bibitem[Göbel et~al.(1994)Göbel, Sander, Schneider, and
  Valencia]{gobel_correlated_1994}
U.~Göbel, C.~Sander, R.~Schneider, and A.~Valencia.
\newblock Correlated mutations and residue contacts in proteins.
\newblock \emph{Proteins}, 18\penalty0 (4):\penalty0 309--317, Apr. 1994.
\newblock ISSN 0887-3585.
\newblock \doi{10.1002/prot.340180402}.

\bibitem[Harvey and Pagel(1991)]{harvey_comparative_1991}
P.~H. Harvey and M.~D. Pagel.
\newblock \emph{The {Comparative} {Method} {In} {Evolutionary} {Biology}}.
\newblock Oxford University Press, U.S.A., Oxford ; New York, June 1991.
\newblock ISBN 978-0-19-854640-5.

\bibitem[Holley et~al.(1965)Holley, Apgar, Everett, Madison, Marquisee,
  Merrill, Penswick, and Zamir]{holley_structure_1965}
R.~W. Holley, J.~Apgar, G.~A. Everett, J.~T. Madison, M.~Marquisee, S.~H.
  Merrill, J.~R. Penswick, and A.~Zamir.
\newblock Structure of a ribonucleic acid.
\newblock \emph{Science}, 147\penalty0 (3664):\penalty0 1462--1465, Mar. 1965.
\newblock ISSN 0036-8075.
\newblock \doi{10.1126/science.147.3664.1462}.

\bibitem[Hopf et~al.(2012)Hopf, Colwell, Sheridan, Rost, Sander, and
  Marks]{hopf_three-dimensional_2012}
T.~A. Hopf, L.~J. Colwell, R.~Sheridan, B.~Rost, C.~Sander, and D.~S. Marks.
\newblock Three-{Dimensional} {Structures} of {Membrane} {Proteins} from
  {Genomic} {Sequencing}.
\newblock \emph{Cell}, 149\penalty0 (7):\penalty0 1607--1621, June 2012.
\newblock ISSN 0092-8674, 1097-4172.
\newblock \doi{10.1016/j.cell.2012.04.012}.
\newblock URL \url{https://www.cell.com/cell/abstract/S0092-8674(12)00509-0}.
\newblock Publisher: Elsevier.

\bibitem[Ignac et~al.(2014)Ignac, Skupin, Sakhanenko, and
  Galas]{ignac_discovering_2014}
T.~M. Ignac, A.~Skupin, N.~A. Sakhanenko, and D.~J. Galas.
\newblock Discovering {Pair}-{Wise} {Genetic} {Interactions}: {An}
  {Information} {Theory}-{Based} {Approach}.
\newblock \emph{PLOS ONE}, 9\penalty0 (3):\penalty0 e92310, Mar. 2014.
\newblock ISSN 1932-6203.
\newblock \doi{10.1371/journal.pone.0092310}.
\newblock URL
  \url{https://journals.plos.org/plosone/article?id=10.1371/journal.pone.0092310}.
\newblock Publisher: Public Library of Science.

\bibitem[Kay et~al.(2005)Kay, Reeves, Olmstead, and Schemske]{kay_rapid_2005}
K.~M. Kay, P.~A. Reeves, R.~G. Olmstead, and D.~W. Schemske.
\newblock Rapid speciation and the evolution of hummingbird pollination in
  neotropical {Costus} subgenus {Costus} ({Costaceae}): evidence from {nrDNA}
  {ITS} and {ETS} sequences.
\newblock \emph{American Journal of Botany}, 92\penalty0 (11):\penalty0
  1899--1910, 2005.
\newblock ISSN 1537-2197.
\newblock \doi{10.3732/ajb.92.11.1899}.
\newblock URL
  \url{https://bsapubs.onlinelibrary.wiley.com/doi/abs/10.3732/ajb.92.11.1899}.
\newblock \_eprint:
  https://bsapubs.onlinelibrary.wiley.com/doi/pdf/10.3732/ajb.92.11.1899.

\bibitem[Kryazhimskiy et~al.(2011)Kryazhimskiy, Dushoff, Bazykin, and
  Plotkin]{kryazhimskiy_prevalence_2011}
S.~Kryazhimskiy, J.~Dushoff, G.~A. Bazykin, and J.~B. Plotkin.
\newblock Prevalence of epistasis in the evolution of influenza {A} surface
  proteins.
\newblock \emph{PLoS Genet.}, 7\penalty0 (2):\penalty0 e1001301, Feb. 2011.
\newblock ISSN 1553-7404.
\newblock \doi{10.1371/journal.pgen.1001301}.

\bibitem[Laine et~al.(2019)Laine, Karami, and Carbone]{laine_gemme_2019}
E.~Laine, Y.~Karami, and A.~Carbone.
\newblock {GEMME}: {A} {Simple} and {Fast} {Global} {Epistatic} {Model}
  {Predicting} {Mutational} {Effects}.
\newblock \emph{Mol Biol Evol}, 36\penalty0 (11):\penalty0 2604--2619, Nov.
  2019.
\newblock ISSN 0737-4038.
\newblock \doi{10.1093/molbev/msz179}.
\newblock URL \url{https://academic.oup.com/mbe/article/36/11/2604/5548199}.
\newblock Publisher: Oxford Academic.

\bibitem[Lande et~al.(1999)Lande, Engen, and Sæther]{lande_spatial_1999}
R.~Lande, S.~Engen, and B.~Sæther.
\newblock Spatial {Scale} of {Population} {Synchrony}: {Environmental}
  {Correlation} versus {Dispersal} and {Density} {Regulation}.
\newblock \emph{The American Naturalist}, 154\penalty0 (3):\penalty0 271--281,
  Sept. 1999.
\newblock ISSN 0003-0147.
\newblock \doi{10.1086/303240}.
\newblock URL \url{https://www.journals.uchicago.edu/doi/full/10.1086/303240}.
\newblock Publisher: The University of Chicago Press.

\bibitem[Lander and Schork(1994)]{lander_genetic_1994}
E.~S. Lander and N.~J. Schork.
\newblock Genetic dissection of complex traits.
\newblock \emph{Science}, 265\penalty0 (5181):\penalty0 2037--2048, Sept. 1994.
\newblock ISSN 0036-8075, 1095-9203.
\newblock \doi{10.1126/science.8091226}.
\newblock URL \url{https://science.sciencemag.org/content/265/5181/2037}.
\newblock Publisher: American Association for the Advancement of Science
  Section: Articles.

\bibitem[Lartillot(2013)]{lartillot_interaction_2013}
N.~Lartillot.
\newblock Interaction between {Selection} and {Biased} {Gene} {Conversion} in
  {Mammalian} {Protein}-{Coding} {Sequence} {Evolution} {Revealed} by a
  {Phylogenetic} {Covariance} {Analysis}.
\newblock \emph{Mol Biol Evol}, 30\penalty0 (2):\penalty0 356--368, Feb. 2013.
\newblock ISSN 0737-4038.
\newblock \doi{10.1093/molbev/mss231}.
\newblock URL \url{https://academic.oup.com/mbe/article/30/2/356/1015523}.
\newblock Publisher: Oxford Academic.

\bibitem[Lartillot and Delsuc(2012)]{lartillot_joint_2012}
N.~Lartillot and F.~Delsuc.
\newblock Joint {Reconstruction} of {Divergence} {Times} and {Life}-{History}
  {Evolution} in {Placental} {Mammals} {Using} a {Phylogenetic} {Covariance}
  {Model}.
\newblock \emph{Evolution}, 66\penalty0 (6):\penalty0 1773--1787, 2012.
\newblock ISSN 1558-5646.
\newblock \doi{10.1111/j.1558-5646.2011.01558.x}.
\newblock URL
  \url{https://onlinelibrary.wiley.com/doi/abs/10.1111/j.1558-5646.2011.01558.x}.
\newblock \_eprint:
  https://onlinelibrary.wiley.com/doi/pdf/10.1111/j.1558-5646.2011.01558.x.

\bibitem[Lartillot and Poujol(2011)]{lartillot_phylogenetic_2011}
N.~Lartillot and R.~Poujol.
\newblock A phylogenetic model for investigating correlated evolution of
  substitution rates and continuous phenotypic characters.
\newblock \emph{Mol. Biol. Evol.}, 28\penalty0 (1):\penalty0 729--744, Jan.
  2011.
\newblock ISSN 1537-1719.
\newblock \doi{10.1093/molbev/msq244}.

\bibitem[Lenormand(2003)]{lenormand_evolution_2003}
T.~Lenormand.
\newblock The evolution of sex dimorphism in recombination.
\newblock \emph{Genetics}, 163\penalty0 (2):\penalty0 811--822, Feb. 2003.
\newblock ISSN 0016-6731.

\bibitem[Lenormand and Dutheil(2005)]{lenormand_recombination_2005}
T.~Lenormand and J.~Dutheil.
\newblock Recombination difference between sexes: a role for haploid selection.
\newblock \emph{PLoS Biol.}, 3\penalty0 (3):\penalty0 e63, Mar. 2005.
\newblock ISSN 1545-7885.
\newblock \doi{10.1371/journal.pbio.0030063}.

\bibitem[Lewontin(2004)]{lewontin_building_2004}
R.~Lewontin.
\newblock Building a science of population biology.
\newblock In R.~S. Singh and M.~K. Uyenoyama, editors, \emph{The {Evolution} of
  {Population} {Biology}}. Cambridge University Press, Jan. 2004.
\newblock ISBN 978-1-139-44954-0.

\bibitem[Long et~al.(2015)Long, Liti, Luptak, and
  Tenaillon]{long_elucidating_2015}
A.~Long, G.~Liti, A.~Luptak, and O.~Tenaillon.
\newblock Elucidating the molecular architecture of adaptation via evolve and
  resequence experiments.
\newblock \emph{Nature Reviews Genetics}, 16\penalty0 (10):\penalty0 567--582,
  Oct. 2015.
\newblock ISSN 1471-0064.
\newblock \doi{10.1038/nrg3937}.
\newblock URL \url{https://www.nature.com/articles/nrg3937}.
\newblock Number: 10 Publisher: Nature Publishing Group.

\bibitem[Maddison(1990)]{maddison_method_1990}
W.~P. Maddison.
\newblock A {Method} for {Testing} the {Correlated} {Evolution} of {Two}
  {Binary} {Characters}: {Are} {Gains} or {Losses} {Concentrated} on {Certain}
  {Branches} of a {Phylogenetic} {Tree}?
\newblock \emph{Evolution}, 44\penalty0 (3):\penalty0 539--557, 1990.
\newblock ISSN 1558-5646.
\newblock \doi{10.1111/j.1558-5646.1990.tb05937.x}.
\newblock URL
  \url{https://onlinelibrary.wiley.com/doi/abs/10.1111/j.1558-5646.1990.tb05937.x}.
\newblock \_eprint:
  https://onlinelibrary.wiley.com/doi/pdf/10.1111/j.1558-5646.1990.tb05937.x.

\bibitem[Mai et~al.(2010)Mai, Sahana, Christiansen, and
  Guldbrandtsen]{mai_genome-wide_2010}
M.~D. Mai, G.~Sahana, F.~B. Christiansen, and B.~Guldbrandtsen.
\newblock A genome-wide association study for milk production traits in
  {Danish} {Jersey} cattle using a {50K} single nucleotide polymorphism chip.
\newblock \emph{J Anim Sci}, 88\penalty0 (11):\penalty0 3522--3528, Nov. 2010.
\newblock ISSN 0021-8812.
\newblock \doi{10.2527/jas.2009-2713}.
\newblock URL \url{https://academic.oup.com/jas/article/88/11/3522/4764149}.
\newblock Publisher: Oxford Academic.

\bibitem[Marks et~al.(2011)Marks, Colwell, Sheridan, Hopf, Pagnani, Zecchina,
  and Sander]{marks_protein_2011}
D.~S. Marks, L.~J. Colwell, R.~Sheridan, T.~A. Hopf, A.~Pagnani, R.~Zecchina,
  and C.~Sander.
\newblock Protein {3D} {Structure} {Computed} from {Evolutionary} {Sequence}
  {Variation}.
\newblock \emph{PLOS ONE}, 6\penalty0 (12):\penalty0 e28766, Dec. 2011.
\newblock ISSN 1932-6203.
\newblock \doi{10.1371/journal.pone.0028766}.
\newblock URL
  \url{https://journals.plos.org/plosone/article?id=10.1371/journal.pone.0028766}.
\newblock Publisher: Public Library of Science.

\bibitem[Martins(1994)]{martins_estimating_1994}
E.~P. Martins.
\newblock Estimating the {Rate} of {Phenotypic} {Evolution} from {Comparative}
  {Data}.
\newblock \emph{The American Naturalist}, 144\penalty0 (2):\penalty0 193--209,
  1994.
\newblock ISSN 0003-0147.
\newblock URL \url{https://www.jstor.org/stable/2463156}.
\newblock Publisher: [University of Chicago Press, American Society of
  Naturalists].

\bibitem[Martins and Hansen(1997)]{martins_phylogenies_1997}
E.~P. Martins and T.~F. Hansen.
\newblock Phylogenies and the {Comparative} {Method}: {A} {General} {Approach}
  to {Incorporating} {Phylogenetic} {Information} into the {Analysis} of
  {Interspecific} {Data}.
\newblock \emph{The American Naturalist}, 149\penalty0 (4):\penalty0 646--667,
  1997.
\newblock ISSN 0003-0147.
\newblock URL \url{https://www.jstor.org/stable/2463542}.
\newblock Publisher: [University of Chicago Press, American Society of
  Naturalists].

\bibitem[McVean(2009)]{mcvean_genealogical_2009}
G.~McVean.
\newblock A {Genealogical} {Interpretation} of {Principal} {Components}
  {Analysis}.
\newblock \emph{PLOS Genetics}, 5\penalty0 (10):\penalty0 e1000686, Oct. 2009.
\newblock ISSN 1553-7404.
\newblock \doi{10.1371/journal.pgen.1000686}.
\newblock URL
  \url{https://journals.plos.org/plosgenetics/article?id=10.1371/journal.pgen.1000686}.
\newblock Publisher: Public Library of Science.

\bibitem[Mendel(1866)]{mendel_hybrid_1866}
J.~G. Mendel.
\newblock Versuche über pflanzenhybriden (eng. experiments in plant
  hybridization).
\newblock \emph{Verhandlungen des naturforschenden Vereines in Brünn (ENG.
  Journal of the Royal Horticultural Society, 1901)}, 26:\penalty0 1--32.,
  1866.

\bibitem[Meyer et~al.(2019)Meyer, Dib, Silvestro, and
  Salamin]{meyer_simultaneous_2019}
X.~Meyer, L.~Dib, D.~Silvestro, and N.~Salamin.
\newblock Simultaneous {Bayesian} inference of phylogeny and molecular
  coevolution.
\newblock \emph{Proc. Natl. Acad. Sci. U.S.A.}, 116\penalty0 (11):\penalty0
  5027--5036, 2019.
\newblock ISSN 1091-6490.
\newblock \doi{10.1073/pnas.1813836116}.

\bibitem[Moore(2001)]{moore_rediscovery_2001}
R.~Moore.
\newblock The "{Rediscovery}" of {Mendel}'s {Work}.
\newblock \emph{Bioscene}, 27\penalty0 (2):\penalty0 13--24, 2001.

\bibitem[Morcos et~al.(2011)Morcos, Pagnani, Lunt, Bertolino, Marks, Sander,
  Zecchina, Onuchic, Hwa, and Weigt]{morcos_direct-coupling_2011}
F.~Morcos, A.~Pagnani, B.~Lunt, A.~Bertolino, D.~S. Marks, C.~Sander,
  R.~Zecchina, J.~N. Onuchic, T.~Hwa, and M.~Weigt.
\newblock Direct-coupling analysis of residue coevolution captures native
  contacts across many protein families.
\newblock \emph{Proc. Natl. Acad. Sci. U.S.A.}, 108\penalty0 (49):\penalty0
  E1293--1301, Dec. 2011.
\newblock ISSN 1091-6490.
\newblock \doi{10.1073/pnas.1111471108}.

\bibitem[Morgan(2018)]{morgan_theory_2018}
T.~H. Morgan.
\newblock \emph{The {Theory} of the {Gene}: {2D} {Ed}}.
\newblock Creative Media Partners, LLC, Oct. 2018.
\newblock ISBN 978-0-343-29036-8.
\newblock Google-Books-ID: dIqDAQAACAAJ.

\bibitem[Morton(1955)]{morton_sequential_1955}
N.~E. Morton.
\newblock Sequential tests for the detection of linkage.
\newblock \emph{Am. J. Hum. Genet.}, 7\penalty0 (3):\penalty0 277--318, Sept.
  1955.
\newblock ISSN 0002-9297.

\bibitem[Neher(1994)]{neher_how_1994}
E.~Neher.
\newblock How frequent are correlated changes in families of protein sequences?
\newblock \emph{Proc. Natl. Acad. Sci. U.S.A.}, 91\penalty0 (1):\penalty0
  98--102, Jan. 1994.
\newblock ISSN 0027-8424.
\newblock \doi{10.1073/pnas.91.1.98}.

\bibitem[Novembre and Stephens(2008)]{novembre_interpreting_2008}
J.~Novembre and M.~Stephens.
\newblock Interpreting principal component analyses of spatial population
  genetic variation.
\newblock \emph{Nature Genetics}, 40\penalty0 (5):\penalty0 646--649, May 2008.
\newblock ISSN 1546-1718.
\newblock \doi{10.1038/ng.139}.
\newblock URL \url{https://www.nature.com/articles/ng.139}.
\newblock Number: 5 Publisher: Nature Publishing Group.

\bibitem[Nyholt(2000)]{nyholt_all_2000}
D.~R. Nyholt.
\newblock All {LODs} {Are} {Not} {Created} {Equal}*.
\newblock \emph{The American Journal of Human Genetics}, 67\penalty0
  (2):\penalty0 282--288, Aug. 2000.
\newblock ISSN 0002-9297, 1537-6605.
\newblock \doi{10.1086/303029}.
\newblock URL \url{https://www.cell.com/ajhg/abstract/S0002-9297(07)62639-1}.
\newblock Publisher: Elsevier.

\bibitem[Ozaki et~al.(2002)Ozaki, Ohnishi, Iida, Sekine, Yamada, Tsunoda, Sato,
  Sato, Hori, Nakamura, and Tanaka]{ozaki_functional_2002}
K.~Ozaki, Y.~Ohnishi, A.~Iida, A.~Sekine, R.~Yamada, T.~Tsunoda, H.~Sato,
  H.~Sato, M.~Hori, Y.~Nakamura, and T.~Tanaka.
\newblock Functional {SNPs} in the lymphotoxin-$\alpha$ gene that are
  associated with susceptibility to myocardial infarction.
\newblock \emph{Nature Genetics}, 32\penalty0 (4):\penalty0 650--654, Dec.
  2002.
\newblock ISSN 1546-1718.
\newblock \doi{10.1038/ng1047}.
\newblock URL \url{https://www.nature.com/articles/ng1047z}.
\newblock Number: 4 Publisher: Nature Publishing Group.

\bibitem[Pagel(1994)]{pagel_detecting_1994}
M.~Pagel.
\newblock Detecting correlated evolution on phylogenies: a general method for
  the comparative analysis of discrete characters.
\newblock \emph{Proceedings of the Royal Society of London. Series B:
  Biological Sciences}, 255\penalty0 (1342):\penalty0 37--45, Jan. 1994.
\newblock \doi{10.1098/rspb.1994.0006}.
\newblock URL
  \url{https://royalsocietypublishing.org/doi/10.1098/rspb.1994.0006}.
\newblock Publisher: Royal Society.

\bibitem[Paradis and Claude(2002)]{paradis_analysis_2002}
E.~Paradis and J.~Claude.
\newblock Analysis of comparative data using generalized estimating equations.
\newblock \emph{J. Theor. Biol.}, 218\penalty0 (2):\penalty0 175--185, Sept.
  2002.
\newblock ISSN 0022-5193.
\newblock \doi{10.1006/jtbi.2002.3066}.

\bibitem[Pazos and Valencia(2001)]{pazos_similarity_2001}
F.~Pazos and A.~Valencia.
\newblock Similarity of phylogenetic trees as indicator of protein-protein
  interaction.
\newblock \emph{Protein Eng.}, 14\penalty0 (9):\penalty0 609--614, Sept. 2001.
\newblock ISSN 0269-2139.
\newblock \doi{10.1093/protein/14.9.609}.

\bibitem[Pearson and Galton(1895)]{pearson_vii_1895}
K.~Pearson and F.~Galton.
\newblock {VII}. {Note} on regression and inheritance in the case of two
  parents.
\newblock \emph{Proceedings of the Royal Society of London}, 58\penalty0
  (347-352):\penalty0 240--242, Jan. 1895.
\newblock \doi{10.1098/rspl.1895.0041}.
\newblock URL
  \url{https://royalsocietypublishing.org/doi/10.1098/rspl.1895.0041}.
\newblock Publisher: Royal Society.

\bibitem[Pensar et~al.(2019)Pensar, Puranen, Arnold, MacAlasdair, Kuronen,
  Tonkin-Hill, Pesonen, Xu, Sipola, Sánchez-Busó, Lees, Chewapreecha,
  Bentley, Harris, Parkhill, Croucher, and Corander]{pensar_genome-wide_2019}
J.~Pensar, S.~Puranen, B.~Arnold, N.~MacAlasdair, J.~Kuronen, G.~Tonkin-Hill,
  M.~Pesonen, Y.~Xu, A.~Sipola, L.~Sánchez-Busó, J.~A. Lees, C.~Chewapreecha,
  S.~D. Bentley, S.~R. Harris, J.~Parkhill, N.~J. Croucher, and J.~Corander.
\newblock Genome-wide epistasis and co-selection study using mutual
  information.
\newblock \emph{Nucleic Acids Res}, 47\penalty0 (18):\penalty0 e112--e112, Oct.
  2019.
\newblock ISSN 0305-1048.
\newblock \doi{10.1093/nar/gkz656}.
\newblock URL \url{https://academic.oup.com/nar/article/47/18/e112/5541093}.
\newblock Publisher: Oxford Academic.

\bibitem[Phillips(2008)]{phillips_epistasis_2008}
P.~C. Phillips.
\newblock Epistasis — the essential role of gene interactions in the
  structure and evolution of genetic systems.
\newblock \emph{Nat Rev Genet}, 9\penalty0 (11):\penalty0 855--867, Nov. 2008.
\newblock ISSN 1471-0064.
\newblock \doi{10.1038/nrg2452}.
\newblock URL \url{https://www.nature.com/articles/nrg2452}.
\newblock Number: 11 Publisher: Nature Publishing Group.

\bibitem[Pollock et~al.(1999)Pollock, Taylor, and
  Goldman]{pollock_coevolving_1999}
D.~D. Pollock, W.~R. Taylor, and N.~Goldman.
\newblock Coevolving protein residues: maximum likelihood identification and
  relationship to structure.
\newblock \emph{J. Mol. Biol.}, 287\penalty0 (1):\penalty0 187--198, Mar. 1999.
\newblock ISSN 0022-2836.
\newblock \doi{10.1006/jmbi.1998.2601}.

\bibitem[Price et~al.(2006)Price, Patterson, Plenge, Weinblatt, Shadick, and
  Reich]{price_principal_2006}
A.~L. Price, N.~J. Patterson, R.~M. Plenge, M.~E. Weinblatt, N.~A. Shadick, and
  D.~Reich.
\newblock Principal components analysis corrects for stratification in
  genome-wide association studies.
\newblock \emph{Nature Genetics}, 38\penalty0 (8):\penalty0 904--909, Aug.
  2006.
\newblock ISSN 1546-1718.
\newblock \doi{10.1038/ng1847}.
\newblock URL \url{https://www.nature.com/articles/ng1847}.
\newblock Number: 8 Publisher: Nature Publishing Group.

\bibitem[Pritchard et~al.(2000)Pritchard, Stephens, Rosenberg, and
  Donnelly]{pritchard_association_2000}
J.~K. Pritchard, M.~Stephens, N.~A. Rosenberg, and P.~Donnelly.
\newblock Association {Mapping} in {Structured} {Populations}.
\newblock \emph{The American Journal of Human Genetics}, 67\penalty0
  (1):\penalty0 170--181, July 2000.
\newblock ISSN 0002-9297.
\newblock \doi{10.1086/302959}.
\newblock URL
  \url{http://www.sciencedirect.com/science/article/pii/S0002929707624422}.

\bibitem[Puranen et~al.(2018)Puranen, Pesonen, Pensar, Xu, Lees, Bentley,
  Croucher, and Corander]{puranen_superdca_2018}
S.~Puranen, M.~Pesonen, J.~Pensar, Y.~Y. Xu, J.~A. Lees, S.~D. Bentley, N.~J.
  Croucher, and J.~Corander.
\newblock {SuperDCA} for genome-wide epistasis analysis.
\newblock \emph{Microbial Genomics,}, 4\penalty0 (6):\penalty0 e000184, 2018.
\newblock ISSN ,.
\newblock \doi{10.1099/mgen.0.000184}.
\newblock URL
  \url{https://www.microbiologyresearch.org/content/journal/mgen/10.1099/mgen.0.000184}.
\newblock Publisher: Microbiology Society,.

\bibitem[Purcell et~al.(2007)Purcell, Neale, Todd-Brown, Thomas, Ferreira,
  Bender, Maller, Sklar, de~Bakker, Daly, and Sham]{purcell_plink_2007}
S.~Purcell, B.~Neale, K.~Todd-Brown, L.~Thomas, M.~A.~R. Ferreira, D.~Bender,
  J.~Maller, P.~Sklar, P.~I.~W. de~Bakker, M.~J. Daly, and P.~C. Sham.
\newblock {PLINK}: {A} {Tool} {Set} for {Whole}-{Genome} {Association} and
  {Population}-{Based} {Linkage} {Analyses}.
\newblock \emph{The American Journal of Human Genetics}, 81\penalty0
  (3):\penalty0 559--575, Sept. 2007.
\newblock ISSN 0002-9297.
\newblock \doi{10.1086/519795}.
\newblock URL
  \url{http://www.sciencedirect.com/science/article/pii/S0002929707613524}.

\bibitem[Rensch(1950)]{Rensch1950DieAD}
B.~Rensch.
\newblock \emph{Die Abh{\"a}ngigkeit der relativen Sexualdifferenz von der
  K{\"o}rpergr{\"o}{\ss}e}, volume~1.
\newblock Bonner zoologische Beiträge, 1950.

\bibitem[Ridley(1983)]{ridley_explanation_1983}
M.~Ridley.
\newblock \emph{The {Explanation} of {Organic} {Diversity}: {The} {Comparative}
  {Method} and {Adaptations} for {Mating}}.
\newblock Oxford University Press, Oxford : Oxford ; New York, Dec. 1983.
\newblock ISBN 978-0-19-857597-9.

\bibitem[Shindyalov et~al.(1994)Shindyalov, Kolchanov, and
  Sander]{shindyalov_can_1994}
I.~N. Shindyalov, N.~A. Kolchanov, and C.~Sander.
\newblock Can three-dimensional contacts in protein structures be predicted by
  analysis of correlated mutations?
\newblock \emph{Protein Eng.}, 7\penalty0 (3):\penalty0 349--358, Mar. 1994.
\newblock ISSN 0269-2139.
\newblock \doi{10.1093/protein/7.3.349}.

\bibitem[Szurmant and Weigt(2018)]{szurmant_inter-residue_2018}
H.~Szurmant and M.~Weigt.
\newblock Inter-residue, inter-protein and inter-family coevolution: bridging
  the scales.
\newblock \emph{Curr. Opin. Struct. Biol.}, 50:\penalty0 26--32, 2018.
\newblock ISSN 1879-033X.
\newblock \doi{10.1016/j.sbi.2017.10.014}.

\bibitem[Talavera et~al.(2015)Talavera, Lovell, and
  Whelan]{talavera_covariation_2015}
D.~Talavera, S.~C. Lovell, and S.~Whelan.
\newblock Covariation {Is} a {Poor} {Measure} of {Molecular} {Coevolution}.
\newblock \emph{Mol. Biol. Evol.}, 32\penalty0 (9):\penalty0 2456--2468, Sept.
  2015.
\newblock ISSN 1537-1719.
\newblock \doi{10.1093/molbev/msv109}.

\bibitem[Tillier and Collins(1998)]{tillier_high_1998}
E.~R. Tillier and R.~A. Collins.
\newblock High apparent rate of simultaneous compensatory base-pair
  substitutions in ribosomal {RNA}.
\newblock \emph{Genetics}, 148\penalty0 (4):\penalty0 1993--2002, Apr. 1998.
\newblock ISSN 0016-6731.

\bibitem[Tillier and Collins(1995)]{tillier_neighbor_1995}
E.~R.~M. Tillier and R.~A. Collins.
\newblock Neighbor {Joining} and {Maximum} {Likelihood} with {RNA} {Sequences}:
  {Addressing} the {Interdependence} of {Sites}.
\newblock \emph{Mol Biol Evol}, 12\penalty0 (1):\penalty0 7--7, Jan. 1995.
\newblock ISSN 0737-4038.
\newblock \doi{10.1093/oxfordjournals.molbev.a040195}.
\newblock URL \url{https://academic.oup.com/mbe/article/12/1/7/999573}.
\newblock Publisher: Oxford Academic.

\bibitem[Tufféry and Darlu(2000)]{tuffery_exploring_2000}
P.~Tufféry and P.~Darlu.
\newblock Exploring a phylogenetic approach for the detection of correlated
  substitutions in proteins.
\newblock \emph{Mol. Biol. Evol.}, 17\penalty0 (11):\penalty0 1753--1759, Nov.
  2000.
\newblock ISSN 0737-4038.
\newblock \doi{10.1093/oxfordjournals.molbev.a026273}.

\bibitem[Van~Valen(1973)]{van_valen_new_1973}
L.~Van~Valen.
\newblock A {New} {Evolutionary} {Law}.
\newblock \emph{Evol. Theory}, 1\penalty0 (1):\penalty0 1--30, July 1973.

\bibitem[Visser and Krug(2014)]{visser_empirical_2014}
J.~A. G. M.~d. Visser and J.~Krug.
\newblock Empirical fitness landscapes and the predictability of evolution.
\newblock \emph{Nat Rev Genet}, 15\penalty0 (7):\penalty0 480--490, July 2014.
\newblock ISSN 1471-0064.
\newblock \doi{10.1038/nrg3744}.
\newblock URL \url{https://www.nature.com/articles/nrg3744}.
\newblock Number: 7 Publisher: Nature Publishing Group.

\bibitem[Weigt et~al.(2009)Weigt, White, Szurmant, Hoch, and
  Hwa]{weigt_identification_2009}
M.~Weigt, R.~A. White, H.~Szurmant, J.~A. Hoch, and T.~Hwa.
\newblock Identification of direct residue contacts in protein–protein
  interaction by message passing.
\newblock \emph{PNAS}, 106\penalty0 (1):\penalty0 67--72, Jan. 2009.
\newblock ISSN 0027-8424, 1091-6490.
\newblock \doi{10.1073/pnas.0805923106}.
\newblock URL \url{https://www.pnas.org/content/106/1/67}.
\newblock Publisher: National Academy of Sciences Section: Physical Sciences.

\bibitem[Westhof and Auffinger(2012)]{westhof_transfer_2012}
E.~Westhof and P.~Auffinger.
\newblock Transfer {RNA} structure.
\newblock In \emph{{eLS}}. American Cancer Society, 2012.
\newblock ISBN 978-0-470-01590-2.
\newblock \doi{10.1002/9780470015902.a0000527.pub2}.
\newblock URL
  \url{https://onlinelibrary.wiley.com/doi/abs/10.1002/9780470015902.a0000527.pub2}.
\newblock \_eprint:
  https://onlinelibrary.wiley.com/doi/pdf/10.1002/9780470015902.a0000527.pub2.

\bibitem[Woolhouse et~al.(2002)Woolhouse, Webster, Domingo, Charlesworth, and
  Levin]{woolhouse_biological_2002}
M.~E.~J. Woolhouse, J.~P. Webster, E.~Domingo, B.~Charlesworth, and B.~R.
  Levin.
\newblock Biological and biomedical implications of the co-evolution of
  pathogens and their hosts.
\newblock \emph{Nat Genet}, 32\penalty0 (4):\penalty0 569--577, Dec. 2002.
\newblock ISSN 1546-1718.
\newblock \doi{10.1038/ng1202-569}.
\newblock URL \url{https://www.nature.com/articles/ng1202-569}.
\newblock Number: 4 Publisher: Nature Publishing Group.

\bibitem[Wright(1932)]{wright_roles_1932}
S.~Wright.
\newblock The roles of mutation, inbreeding, crossbreeding and selection in
  evolution.
\newblock In \emph{Proceedings of the Sixth International Congress on
  Genetics}, volume~1, page 356–366, 1932.

\bibitem[Yang(2006)]{yang_computational_2006}
Z.~Yang.
\newblock \emph{Computational {Molecular} {Evolution}}.
\newblock Oxford University Press, Oxford, Dec. 2006.
\newblock ISBN 978-0-19-856702-8.

\bibitem[Yeang(2008)]{yeang_identifying_2008}
C.-H. Yeang.
\newblock Identifying coevolving partners from paralogous gene families.
\newblock \emph{Evol. Bioinform. Online}, 4:\penalty0 97--107, Apr. 2008.
\newblock ISSN 1176-9343.
\newblock \doi{10.4137/ebo.s621}.

\bibitem[Yeang and Haussler(2007)]{yeang_detecting_2007}
C.-H. Yeang and D.~Haussler.
\newblock Detecting coevolution in and among protein domains.
\newblock \emph{PLoS Comput. Biol.}, 3\penalty0 (11):\penalty0 e211, Nov. 2007.
\newblock ISSN 1553-7358.
\newblock \doi{10.1371/journal.pcbi.0030211}.

\bibitem[Yeang et~al.(2007)Yeang, Darot, Noller, and
  Haussler]{yeang_detecting_rna_2007}
C.-H. Yeang, J.~F.~J. Darot, H.~F. Noller, and D.~Haussler.
\newblock Detecting the coevolution of biosequences--an example of {RNA}
  interaction prediction.
\newblock \emph{Mol. Biol. Evol.}, 24\penalty0 (9):\penalty0 2119--2131, Sept.
  2007.
\newblock ISSN 0737-4038.
\newblock \doi{10.1093/molbev/msm142}.

\bibitem[Yi and Dean(2019)]{yi_adaptive_2019}
X.~Yi and A.~M. Dean.
\newblock Adaptive {Landscapes} in the {Age} of {Synthetic} {Biology}.
\newblock \emph{Mol Biol Evol}, 36\penalty0 (5):\penalty0 890--907, May 2019.
\newblock ISSN 0737-4038.
\newblock \doi{10.1093/molbev/msz004}.
\newblock URL \url{https://academic.oup.com/mbe/article/36/5/890/5290102}.
\newblock Publisher: Oxford Academic.

\end{thebibliography}
\bibliographystyle{abbrvnat}

\end{document}